\documentclass[11pt]{article}
\usepackage[text={6in,8.5in},centering]{geometry}
\usepackage{graphicx}
\usepackage{amssymb}
\usepackage{epstopdf}
\usepackage{natbib}

\usepackage{url}
\usepackage{multirow}
\usepackage{amsthm}
\usepackage[symbol,perpage]{footmisc}  %\usepackage[multiple]{footmisc}

\newcommand{\captionfonts}{\small}
\makeatletter  % Allow the use of @ in command names
\long\def\@makecaption#1#2{%
  \vskip\abovecaptionskip
  \sbox\@tempboxa{{\captionfonts #1: #2}}%
  \ifdim \wd\@tempboxa >\hsize
    {\captionfonts #1: #2\par}
  \else
    \hbox to\hsize{\hfil\box\@tempboxa\hfil}%
  \fi
  \vskip\belowcaptionskip}
\makeatother   % Cancel the effect of \makeatletter

\newcommand{\abs}[1]{|#1|}
\newcommand{\argmin}{\mathop{\mathrm{argmin}}\nolimits}

\newcommand{\eps}{\varepsilon}

\newcommand{\upstep}{\ensuremath{\uparrow}}
\newcommand{\smalldownstep}{\ensuremath{\downarrow}}
\newcommand{\bigdownstep}{\ensuremath{{\downarrow \downarrow}}}

\newtheorem{theorem}{Theorem}[section]
\newtheorem*{theorem*}{Theorem}

\newtheorem{lemma}{Lemma}[section]
\newtheorem*{lemma*}{Lemma}
\newtheorem{corollary}{Corollary}[section]
\newtheorem{conjecture}{Conjecture}[section]

\newtheorem*{claim*}{Claim}

\title{Robust Stochastic Chemical Reaction Networks\\and Bounded Tau-Leaping}
\author{David Soloveichik\footnote{Department of CNS, California Institute of Technology,
Mail Code 136-93, Pasadena, CA 91125-9300, USA.  Voice: (626) 395-5707, Fax: (626) 584-0630}
\\ 
{\tt dsolov@caltech.edu} \\
}
\date{}                                           % Activate to display a given date or no date

\begin{document}
\maketitle

\begin{abstract}
The behavior of some stochastic chemical reaction networks is largely unaffected by slight inaccuracies in reaction rates. We formalize the robustness of state probabilities to reaction rate deviations, and describe a formal connection between robustness and efficiency of simulation. Without robustness guarantees, stochastic simulation seems to require computational time proportional to the total number of reaction events. Even if the concentration (molecular count per volume) stays bounded, the number of reaction events can be linear in the duration of simulated time and total molecular count. We show that the behavior of robust systems can be predicted such that the computational work scales linearly with the duration of simulated time and concentration, and only polylogarithmically in the total molecular count. Thus our asymptotic analysis captures the dramatic speedup when molecular counts are large, and shows that for bounded concentrations the computation time is essentially invariant with molecular count. Finally, by noticing that even robust stochastic chemical reaction networks are capable of embedding complex computational problems, we argue that the linear dependence on simulated time and concentration is likely optimal.
\end{abstract}

\section{Introduction}

The stochastic chemical reaction network (SCRN) model of chemical kinetics 
is used in chemistry, physics, and computational biology.
It describes interactions involving integer number of molecules as Markov jump processes~\citep{mcquarrie67,vankampen97,erdi89,gillespie92},
and is used in domains where the traditional model of deterministic continuous mass action kinetics is invalid due to small molecular counts.
Small molecular counts are prevalent in biology: 
for example, over 80\% of the genes in the \textit{E. coli} chromosome are expressed at fewer than a hundred copies per cell, with some key control factors present in quantities under a dozen~\citep{Guptasarma95,Levin99}.
Indeed, experimental observations and computer simulations have confirmed that stochastic effects can be physiologically significant~\citep{McAdams97,Elowitz02a,elowitz06}.
%Accurate simulation of many biochemical processes seems to require explicitly tracking the integer molecular counts of each chemical species according to the kinetics described below.  
Consequently, the stochastic model is widely employed for modeling cellular processes (e.g.,\ \citet{Arkin98}) and is included in numerous software packages~\citep{Vasudeva04,STOCKS,BioNetS}.\footnote{Some stochastic simulation implementations on the web: Systems Biology Workbench: \url{http://sbw.sourceforge.net}; BioSpice: \url{http://biospice.lbl.gov}; Stochastirator: \url{http://opnsrcbio.molsci.org}; STOCKS: \url{http://www.sysbio.pl/stocks}; BioNetS: \url{http://x.amath.unc.edu:16080/BioNetS}; SimBiology package for {MATLAB}: \url{http://www.mathworks.com/products/simbiology/index.html}}
The stochastic model becomes equivalent to the classical law of mass action when the molecular counts of all participating species are large~\citep{kurtz72,ethier86}.

Gillespie's stochastic simulation algorithm (SSA) can be used to model the behavior of SCRNs~\citep{Gillespie77}.
However, simulation of systems of interest often requires an unfeasible amount of computational time.
Some work has focused on optimizing simulation of large SCRNs --- many different species and reaction channels.  
For example, certain tricks can improve the speed of deciding which reaction occurs next if there are many possible choices (e.g.,~\citet{Gibson00}).  However, for the purposes of this paper we suppose that the number of species and reactions is relatively small, and that it is fundamentally the number of reaction occurrences in a given interval of time that presents the difficulty.
Because SSA simulates every single reaction event,
simulation is slow when the number of reaction events is large.
%Simulation can be slow because of the large number of reaction events that occur in the interval of simulated time.

On the face of it, simulation should be possible without explicitly modeling every reaction occurrence.
In the mass action limit, fast simulation is achieved using numerical ODE solvers.
The complexity of the simulation does not scale at all with the actual number of reaction occurrences but with overall simulation time and the concentration of the species.
If the volume gets larger without a significant increase in concentration, 
mass action ODE solvers achieve a profound difference in computation time compared to SSA.\footnote{
As an illustrative example, a prokaryotic cell and a eukaryotic cell may have similar concentrations of proteins but vastly different volumes.}
Moreover maximum concentration is essentially always bounded,
because the model is only valid for solutions dilute enough to be well mixed,
and ultimately because of the finite density of matter.
However, mass action simulation can only be applied if molecular counts of \emph{all} the species are large.
Even one species that maintains a low molecular count and interacts with other species prevents the use of mass action ODE solvers.

Another reason why it seems that it should be possible to simulate stochastic chemical systems quickly,
is that for many systems the behavior of interest does not depend crucially upon details of events.
For example biochemical networks tend to be robust to variations in concentrations and kinetic parameters~\citep{morohashi2002rmp,alon2007isb}.
If these systems are robust to many kinds of perturbations, including sloppiness in simulation,
can we take advantage of this to speed up simulation?
For example, can we approach the speed of ODEs but allow molecular counts of some species to be small?
%First we need to specify exactly what kind of robustness these systems have that may be useful to us.
Indeed, tau-leaping algorithms (e.g.,~\citet{Gillespie01,rathinam2003ssc,gillespie06}, see~\citet{gillespie07} for a review) are based on the idea that if we allow reaction propensities to remain constant for some amount of time $\tau$, 
but therefore deviate slightly from their correct values, we don't have to explicitly simulate every reaction that occurs in this period of time (and can thus ``leap" by amount of time $\tau$).

In this paper we formally define robustness of the probability that the system is in a certain state at a certain time to perturbations in reaction propensities.
We also provide a method for proving that certain simple systems are robust.
We then describe a new approximate stochastic simulation algorithm called bounded tau-leaping (BTL),
which naturally follows from our definition of robustness,
and provably provides correct answers for robust systems.
In contrast to Gillespie's and others' versions of tau-leaping, in each step of our algorithm the leap time, rather than being a function of the current state, is a random variable.
This algorithm naturally avoids some pitfalls of tau-leaping:
the concentrations cannot become negative, 
and the algorithm scales to SSA when necessary,
in a way that there is always at least one reaction per leap.
However, in the cases when there are ``opposing reactions" (canceling or partially cancelling each other) other forms of tau-leaping may be significantly faster (e.g.,~\citet{rathinam2007rem}).

BTL seems more amenable to theoretical analysis than Gillespie's versions~\citep{Gillespie01,gillespie2003ils,gillespie06},
and may thus act as a stand-in for approximate simulation algorithms in analytic investigations.
%Indeed, here we derive an analytic upper-bound the number of leaps it takes.
In this paper we use the language and tools of computational complexity theory 
to formally study how the number of leaps that BTL takes varies with the maximum molecular count $m$, time span of the simulation $t$, and volume $V$.
In line with the basic computational complexity paradigm,
our analysis is asymptotic and worst-case.
``Asymptotic" means that we do not evaluate the exact number of leaps but rather look at the functional form of the dependence of their number on $m$, $t$, and $V$.
This is easier to derive and allows for making fundamental distinctions (e.g.,\ an exponential function is fundamentally larger than a polynomial function) without getting lost in the details.
``Worst-case" means that we will not study the behavior of our algorithm on any particular chemical system but rather upper-bound the number of leaps our algorithm takes independent of the chemical system.
This will allow us to know that no matter what the system we are trying to simulate, 
it will not be worse than our bound.

In this computational complexity paradigm, we show that indeed robustness helps.
We prove an upper bound on the number of steps our algorithm takes that is logarithmic in $m$, and linear in $t$ and total concentration $C = m/V$.
This can be contrasted with the exact SSA algorithm which, in the worst case, takes a number of steps that is linear in $m$, $t$, and $C$.
Since a logarithmic dependence is much smaller than a linear one, 
BTL is provably ``closer" to the speed of ODE solvers for mass action systems which have no dependence on $m$.\footnote{
Indeed, the total molecular count $m$ can be extremely large compared to its logarithm ---
e.g., Avogadro's number $= 6 \times 10^{23}$ while its $\log_2$ is only $79$.}

Finally we ask whether it is possible to improve upon BTL for robust systems,
or did we exhaust the speed gains that can be obtained due to robustness?  
In the last section of the paper we connect this question to a conjecture in computer science that is believed to be true.
With this conjecture we prove that there are robust systems whose behavior cannot be predicted in fewer computational steps than the number of leaps that BTL makes, 
ignoring multiplicative constant factors and powers of $\log{m}$.
We believe other versions of tau-leaping have similar worst-case complexities as our algorithm,
but proving equivalent results for them remains open.

%To focus on this source of complexity we consider what happens to the computation speed of our algorithms as the simulation time and the size of the ``test tube" in terms of volume and the number of molecules varies.
%The chemical reactions, rate constants, and other parameters (such as approximation parameters which we'll encounter later) are assumed to be fixed.

\section{Model and Definitions}   \label{sec:definitions}

A \emph{Stochastic Chemical Reaction Network} (SCRN) $\mathcal{S}$ specifies a set of $N$ \emph{species} $S_i$ $(i \in \{1, \dots, N\})$ and $M$ \emph{reactions} $R_j$ $(j \in \{1, \dots, M\})$.
The \emph{state} of $\mathcal{S}$ is a vector $\vec{x} \in \mathbb{N}^N$ indicating the integral molecular counts of the species.\footnote{$\mathbb{N} = \{0,1,2,\dots\}$ and $\mathbb{Z} = \{\dots , -1, 0,1,\dots\}$.}
A reaction $R_j$ specifies a reactants' stoichiometry vector $\vec{r_j} \in \mathbb{N}^N$, a products' stoichiometry vector $\vec{p_j} \in \mathbb{N}^N$, and a real-valued rate constant $k_j > 0$.
We describe reaction stoichiometry using a standard chemical ``arrow" notation;
for example, if there are three species, the reaction $R_j$:  $S_1 + S_2 \rightarrow S_1 + 2 S_3$ has reactants vector $\vec{r_j} = (-1, -1, 0)$ and products vector $\vec{p_j} = (1, 0, 2)$.
A reaction $R_j$ is \emph{possible} in state $\vec{x}$ if there are enough reactant molecules: $(\forall i) \; x_i - r_{ij} \geq 0$.
Then if reaction $R_j$ occurs (or ``fires") in state $\vec{x}$, 
the state changes to $\vec{x} + \vec{\nu_j}$, where $\vec{\nu_j} \in \mathbb{Z}^N$ is the state change vector for reaction $R_j$ defined as $\vec{\nu_j} = \vec{p_j} - \vec{r_j}$.
We follow Gillespie and others and allow unary ($S_i \rightarrow \dots$) and bimolecular ($2S_i \rightarrow \dots$ or $S_i + S_{i'} \rightarrow \dots$, $i \neq i'$) reactions only.
Sometimes the model is extended to higher-order reactions~\citep{vankampen97}, but the merit of this is a matter of some controversy.  
%XXX What about reaction:   -> X whose propensity scales proportionally with the volume?

Let us fix an SCRN $\mathcal{S}$.
Given a starting state $\vec{x_0}$ and a fixed volume $V$, we can define a continuous-time Markov process we call an \emph{SSA process}\footnote{It is exactly the stochastic process simulated by Gillespie's  Stochastic Simulation Algorithm (SSA)~\citep{Gillespie77}.} $\mathcal{C}$ of $\mathcal{S}$ according to the following stochastic kinetics.
Given a current state $\vec{x}$, the propensity function $a_j$ of reaction $R_j$ is defined so that 
$a_j(\vec{x}) dt$ is the probability that one $R_j$ reaction will occur in the next infinitesimal time interval $[t, t+dt)$.
If $R_j$ is a unimolecular reaction $S_i \rightarrow \dots$
then the propensity is proportional to the number of molecules of $S_i$ currently present since each is equally likely to react in the next time instant;
specifically, $a_j(\vec{x}) = k_j x_i$ for reaction rate constant $k_j$.
If $R_j$ is a bimolecular reaction $S_i + S_{i'} \rightarrow \dots$, where $i \neq i'$, 
then the reaction propensity is proportional to $x_i x_{i'}$, which is the number of ways of choosing a molecule of $S_i$ and a molecule of $S_{i'}$,  since each pair is equally likely to react in the next time instant. 
Further, the probability that a particular pair reacts in the next time instant is inversely proportional to the volume, resulting in the propensity function $a_j(\vec{x}) = k_j \frac{x_i x_{i'}}{V}$.
If $R_j$ is a bimolecular reaction $2 S_i \rightarrow \dots$
then the number of ways of choosing two molecules of $S_i$ to react is $\frac{x_i (x_i - 1)}{2}$,
and the propensity function is $a_j(\vec{x}) = k_j \frac{x_i (x_i - 1)}{2V}$.

Since the propensity function $a_j$ of reaction $R_j$ is defined so that 
$a_j(\vec{x}) dt$ is the probability that one $R_j$ reaction will occur in the next infinitesimal time interval $[t, t+dt)$, state transitions in the SSA process are equivalently described as follows:
If the system is in state $\vec{x}$, 
no further reactions are possible if $\sum a_j(\vec{x}) = 0$.
Otherwise,
the time until the next reaction occurs is an exponential random variable with rate $\sum_{j} \alpha_j(\vec{x})$.
The probability that next reaction will be a particular $R_{j^*}$ is 
$\alpha_{j^*}(\vec{x}) / \sum_{j} \alpha_j(\vec{x})$.

We are interested in predicting the behavior of SSA processes.
While there are potentially many different questions that we could be trying to answer,
for simplicity we define the \emph{prediction problem} as follows.
%\begin{quote}
Given an SSA process $\mathcal{C}$, a time $t$, a state $\vec{x}$, and $\delta \geq 0$,
predict\footnote{We phrase the prediction problem in terms appropriate for a simulation algorithm.  An alternative formulation would be the problem of estimating the probability that the SSA process is in $\vec{x}$ at time $t$.
To be able to solve this problem using a simulation algorithm we can at most require that with probability at least $\delta_1$ the estimate is within $\delta_2$ of the true probability for some constants $\delta_1, \delta_2 > 0$.
This can be attained by running the simulation algorithm a constant number of times.}
whether $\mathcal{C}$ is in $\vec{x}$ at time $t$,
such that the probability that the prediction is incorrect is at most $\delta$.
%\end{quote}
In other words we are interested in algorithmically generating values of a Bernoulli random variable $I(\vec{x},t)$ such that the probability that $I(\vec{x},t) = 1$ when $\mathcal{C}$ is not in $\vec{x}$ at time $t$ plus the probability that $I(\vec{x},t) = 0$ when $\mathcal{C}$ is in $\vec{x}$ at time $t$ is at most $\delta$.
We assume $\delta$ is some small positive constant.
We can easily extend the prediction problem to a set of states $\Gamma$ rather than a single target state $\vec{x}$ by asking to predict whether the process is in any of the states in $\Gamma$ at time $t$.
Since $\Gamma$ is meant to capture some qualitative feature of the SSA process that is of interest to us, 
it is called an \emph{outcome}.

By decreasing the volume $V$  (which speeds up all bimolecular reactions), increasing $t$, or allowing for more molecules (up to some bound $m$) we are increasing the number of reaction occurrences that we may need to consider.
Thus for a fixed SCRN, one can try to upper bound the computational complexity of the prediction problem as a function of $V$, $t$, and $m$.
Given a molecular count bound $m$, we define the \emph{bounded-count prediction problem} as before, but allowing an arbitrary answer if the molecular count exceeds $m$ within time $t$.
Suppose $\mathcal{P}$ is a bounded-count prediction problem with molecular count bound $m$, error bound $\delta$, about time $t$ and an SSA process in which the volume is $V$.
We then say $\mathcal{P}$ is a \emph{$(m,t,C,\delta)$-prediction problem} where $C = m/V$ is a bound on the maximum concentration.\footnote{Maximum concentration $C$ is a more natural measure of complexity compared to $V$ because similar to $m$ and $t$, computational complexity increases as $C$ increases.}
Fixing some small $\delta$, we study how the computational complexity of solving $(m,t,C,\delta)$-prediction problems may scale with increasing $m$, $t$, and $C$.
If the $(m,t,C,\delta)$-prediction problem is regarding an outcome $\Gamma$ consisting of multiple states, we require the problem of deciding whether a particular state is in $\Gamma$ to be easily solvable.
%\footnote{An extreme example of a prohibited outcome $\Gamma$ is one that is uncomputable.}
Specifically we require it to be solvable in time at most polylogarithmic in $m$,
which is true for any natural problem.

It has been observed that permitting propensities to deviate slightly from their correct values, 
allows for much faster simulation, especially if the molecular counts of some species are large.
This idea forms the basis of approximate stochastic simulation algorithms such as tau-leaping~\citep{Gillespie01}.
As opposed to the exact SSA process described above, 
consider letting the propensity function vary stochastically.
Specifically, we define new propensity functions $a'_j(\vec{x},t) = \xi_j(t) a_j(\vec{x})$ 
where $\{\xi_j(t)\}$ are random variables indexed by reaction and time.
The value of $\xi_j(t)$ describes the deviation from the correct propensity of reaction $R_j$ at time $t$, and should be close to $1$.
For any SSA process $\mathcal{P}$ we can define a new stochastic process called a \emph{perturbation} of $\mathcal{P}$  through the choice of the distributions of $\{\xi_j(t)\}$.
Note that the new process may not be Markov, and may not possess Poisson transition probabilities.
If there is a $0 < \rho < 1$ such that $\forall j,t$,   $(1-\rho) \leq \xi_j(t) \leq (1+\rho)$, then we call the new process a \emph{$\rho$-perturbation}.
There may be systems exhibiting behavior such that any slight inexactness in the calculation of propensities quickly gets amplified and results in qualitatively different behavior.
However, for some processes, if $\rho$ is a small constant, the $\rho$-perturbation may be a good approximation of the SSA process.
That a $\rho$-perturbation is bounded multiplicatively (i.e., that $\xi_j(t)$ acts multiplicatively) 
corresponds to our intuitive notion that proportionally larger deviations are required to have an effect if the affected propensity is large.

We now define our notion of robustness.  
Intuitively, we want the prediction problem to not be affected even if reaction propensities vary slightly.
Formally, we say an SSA process \emph{$\mathcal{C}$ is $(\rho, \delta)$-robust} with respect to state $\vec{x}$ at time $t$ if for any $\rho$-deviating process $\tilde\mathcal{C}$ based on $\mathcal{C}$,
the probability of being in $\vec{x}$ at time $t$ is within plus or minus $\delta$ of the corresponding probability for $\mathcal{C}$.
This definition can be extended to an outcome $\Gamma$ similar to the definition on the prediction problem.
Finally we say an SSA process $\mathcal{C}$ is \emph{$(\rho, \delta)$-robust with respect to a prediction problem} (or bounded-count prediction problem) $\mathcal{P}$ if $\mathcal{C}$ is $(\rho, \delta)$-robust with respect to the same state (or outcome) as specified in $\mathcal{P}$, at the same time $t$ as specified in $\mathcal{P}$.

For simplicity, we often use asymptotic notation.
The notation $O(1)$ is used to denote an unspecified positive constant.
This constant is potentially different every time the expression $O(1)$ appears.

\section{Robustness Examples}      \label{sec:examples}
In this section we elucidate our notion of robustness by considering some examples.
In general, the question of whether a given SSA process is $(\rho, \delta)$-robust for a particular outcome seems a difficult one.
The problem is especially hard because we have to consider every possible $\rho$-perturbation --- thus we may not even be able to give an approximate characterization of robustness by simulation with SSA.
However, we can characterize the robustness of certain (simple) systems.

\begin{figure}[t]
\centering
\scalebox{0.96}{\includegraphics{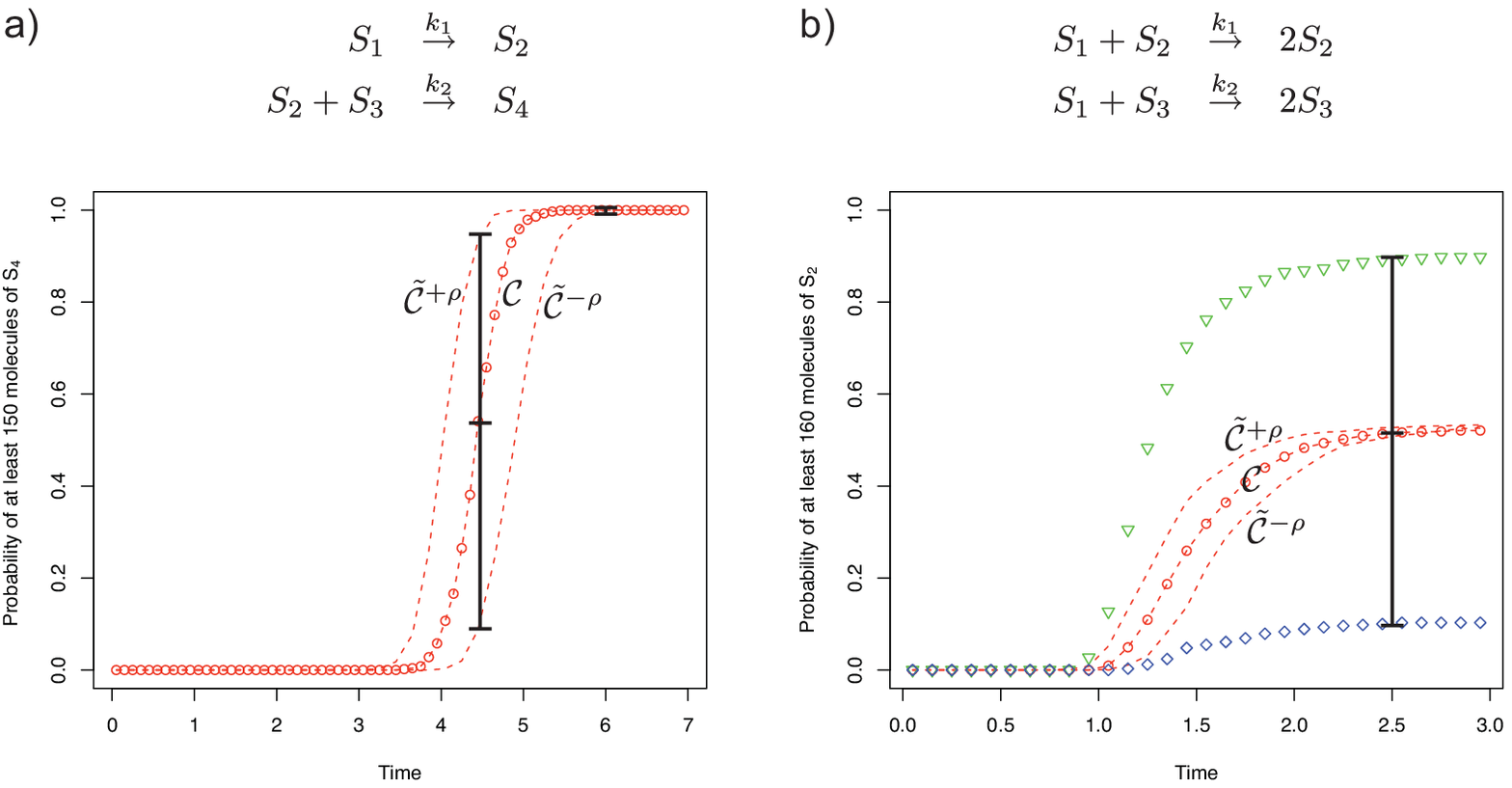}}
\caption{Examples of SCRNs exhibiting contrasting degrees of robustness.
The SSA process $\mathcal{C}$ and outcome $\Gamma$ are defined for the two systems by: 
(a) 
Rate constants: $k_1 = 1$, $k_2 = 0.001$;
start state: $\vec{x_0} = (300,0,300,0)$;
outcome $\Gamma$: $x_4 \geq 150$.
(b) 
Rate constants: $k_1 = 0.01$, $k_2 = 0.01$;
start state: $\vec{x_0} = (300,10,10)$;
outcome $\Gamma$: $x_2 \geq 160$.
Plots show $F^\Gamma(\cdot, t)$ for an SSA process or $\rho$-perturbation estimated from $10^3$ SSA runs.
(Dashed line with circles) Original SSA process $\mathcal{C}$.
(Dashed lines without circles) The two extremal $\rho$-perturbations:
$\tilde{\mathcal{C}}^{+\rho}$ with constant
$\xi_j(t) = 1+\rho$, 
and 
$\tilde{\mathcal{C}}^{-\rho}$ with constant  
$\xi_j(t) = 1-\rho$.
For SCRN (b) we also plot $F^\Gamma(\cdot,t)$ for 
a $\rho$-perturbation with constant $\xi_1(t) = 1+\rho$, $\xi_2(t) = 1-\rho$ (triangles), 
or constant $\xi_1(t) = 1-\rho$, $\xi_2(t) = 1+\rho$ (diamonds).
Perturbation parameter $\rho = 0.1$ throughout.
}
\label{fig:robustness_examples}
\end{figure}

For an SSA process or $\rho$-perturbation $\mathcal{C}$, and outcome $\Gamma$, 
let $F^\Gamma(\mathcal{C},t)$ be the probability of being in $\Gamma$ at time $t$. 
Consider the SCRN shown in Figure~\ref{fig:robustness_examples}(a).
We start with $300$ molecules of $S_1$ and $S_3$ each,
and are interested in the outcome $\Gamma$ of having at least $150$ molecules of $S_4$.
The dashed line with circles shows $F$ for the correct SSA process $\mathcal{C}$.
(All plots of $F$ are estimated from $10^3$ SSA runs.)
The two dashed lines without circles show $F$ for two ``extremal" $\rho$-perturbations:
$\tilde{\mathcal{C}}^{+\rho}$ with constant
$\xi_j(t) = 1+\rho$, 
and 
$\tilde{\mathcal{C}}^{-\rho}$ with constant  
$\xi_j(t) = 1-\rho$.
What can we say about other $\rho$-perturbations, particularly where the $\xi_j(t)$ have much more complicated distributions?
It turns out that for this SCRN and $\Gamma$, 
we can prove that any $\rho$-perturbation falls within the bounds set by the two extremal $\rho$-perturbations $\tilde{\mathcal{C}}^{-\rho}$ and $\tilde{\mathcal{C}}^{+\rho}$.
Thus $F$ for any $\rho$-perturbation falls within the dashed lines.
Formally, $\mathcal{C}$ is monotonic with respect to $\Gamma$ using the definition of monotonicity in Appendix~\ref{sec:monotonicity}.
This is easily proven by Lemma~\ref{lem:syntacticmonotonicity} because every species is a reactant in at most one reaction.
Then by Lemma~\ref{lem:monotonicity}, 
$F^\Gamma(\tilde{\mathcal{C}}^{-\rho}, t) \leq F^\Gamma(\tilde\mathcal{C},t) \leq F^\Gamma(\tilde{\mathcal{C}}^{+\rho}, t)$
for any $\rho$-perturbation $\tilde\mathcal{C}$.

To see how the robustness of this system can be quantified using our definition of $(\rho,\delta)$-robustness,
first consider two time points $t = 4.5$ and $t=6$.
At $t=4.5$, the probability that the correct SSA process $\mathcal{C}$ has produced at least $150$ molecules of $S_4$ is slightly more than $0.5$.
The corresponding probability for $\rho$-perturbations of $\mathcal{C}$ can be no larger than about $0.95$ and no smaller than about $0.1$.
Thus $\mathcal{C}$ is $(\rho,\delta)$-robust with respect to outcome $\Gamma$ at time $t=4.5$ for $\rho = 0.1$ and $\delta$ approximately $0.45$,
but not for smaller $\delta$.
On the other hand at $t=6$,
the dashed lines are essentially on top of each other,
resulting in a tiny $\delta$.
In fact $\delta$ is small for all times less than approximately $3.5$ or greater than approximately $5.5$.

What information did we need to be able to measure $(\rho, \delta)$-robustness?
Processes $\tilde{\mathcal{C}}^{-\rho}$ and $\tilde{\mathcal{C}}^{+\rho}$ are simply $\mathcal{C}$ scaled in time.
Thus knowing how $F^\Gamma(\mathcal{C},t)$ varies with $t$ allows one to quantify $(\rho,\delta)$-robustness at the various times;
$F^\Gamma(\mathcal{C},t)$ can be estimated from multiple SSA runs of $\mathcal{C}$ as in Figure~\ref{fig:robustness_examples}.
Intuitively, 
$\mathcal{C}$ is $(\rho,\delta)$-robust for small $\delta$ at all times $t$ when
$F^\Gamma(\mathcal{C}, t)$ does not change quickly with $t$ (see Appendix~\ref{sec:monotonicity}).
For systems that are not monotonic, knowing how $F^\Gamma(\mathcal{C}, t)$ varies with time may not help with evaluating $(\rho, \delta)$-robustness.

Indeed, for a contrasting example, consider the SCRN in Figure~\ref{fig:robustness_examples}(b).
We start with $300$ molecules of $S_1$, $10$ molecules of $S_2$, and $10$ molecules of $S_3$,
and we are interested in the outcome of having at least $160$ molecules of $S_2$.
Since $S_1$ is a reactant in both reactions, Lemma~\ref{lem:syntacticmonotonicity} cannot be used.
In fact, the figure shows two $\rho$-perturbations (triangles and diamonds) that clearly escape from the boundaries set by the dashed lines.
The triangles show $F$ for the $\rho$-perturbation where the first reaction is maximally sped up and the second reaction is maximally slowed down.
(Vice versa for the diamonds.)
For characterization of the robustness of this system via $(\rho, \delta)$-robustness,
consider the time point $t=2.5$.
The probability of having at least $160$ molecules of $S_2$ in the correct SSA process $\mathcal{C}$ is around $0.5$.
However, this probability for $\rho$-perturbations of $\mathcal{C}$ can deviate by at least approximately $0.4$ upward and downward as seen by the two $\rho$-perturbations (triangles and diamonds).
Thus at this time the system is not $(\rho,\delta)$-robust for $\delta$ approximately $0.4$.
What about other $\rho$-deviations?
It turns out that for this particular system, the two $\rho$-perturbations corresponding to the triangles and diamonds bound $F$ in the same way that $\tilde{\mathcal{C}}^{-\rho}$ and $\tilde{\mathcal{C}}^{+\rho}$ bounded $F$ in the first example (exercise left to the reader).
Nonetheless, for general systems that are not monotonic it is not clear how one can find such bounding $\rho$-perturbation and in fact they likely would not exist.

Of course, there are other types of SSA process that are not like either of the above examples:
e.g.,\ systems that are robust at many times but not monotonic.
General ways of evaluating robustness of such systems remains an important open problem.
%XXX: Say: There are other kind of systems.  Ones that are "actively" robust, such as bistable systems.

Finally, it is important to note that quantifying the robustness of SSA processes, even monotonic ones, 
seems to require computing many SSA runs.
This is self-defeating when in practice one wants to show that the given SSA process is $(\rho, \delta)$-robust in order to justify the use of an approximate simulation algorithm to quickly simulate it.
%solve the corresponding prediction problem with $\delta$ error.
%Nevertheless, as theoretical notions, $\rho$-perturbations and robustness 
%seem to capture something natural,
%something that has remained implicit in previous work on approximate simulation algorithms,
%and further they directly lead to our definition of BTL.
In these cases, we have to consider $(\rho, \delta)$-robustness a theoretical notion only.
Note, however, that it may be much easier to show that a system is \emph{not} robust by comparing the simulation runs of different $\rho$-perturbations, 
since the runs can be quickly obtained using fast approximate simulation algorithms such as that presented in the next section.

\section{Bounded Tau-Leaping}   

\subsection{The Algorithm}     \label{sec:algorithm} 
We argued in the Introduction that sloppiness can allow for faster simulation.
In this section we give a new approximate stochastic simulation algorithm called \emph{bounded tau-leaping} (BTL) that simulates exactly a certain $\rho$-perturbation rather than the original SSA process.
Consequently, 
the algorithm solves the prediction problem with allowed error $\delta$ for $(\rho, \delta)$-robust SSA processes.

The algorithm is a variant of existing tau-leaping algorithms~\citep{gillespie07}.
However, while other tau-leaping algorithms have an implicit notion of robustness,
BTL is formally compatible with our explicit definition.
As we'll see below, our algorithm also has certain other advantages over many previous tau-leaping implementations: it naturally disallows negative concentrations and scales to SSA in a manner that there is always at least one reaction per leap.
It also seems easier to analyze formally;
obtaining a result similar to Theorem~\ref{thm:algspeed} is an open question for other tau-leaping variants.

BTL has overall form typical of tau-leaping algorithms.
Rather than simulating every reaction occurrence explicitly as per the SSA, 
BTL divides the simulation into leaps which group multiple reaction events.
The propensities of all of the reactions are assumed to be fixed throughout the leap.
This is obviously an approximation since each reaction event affects molecular counts and therefore the propensities.
However, this approximation is useful because simulating the system with the assumption that propensities are fixed turns out to be much easier.
Instead of having to draw random variables for each reaction occurrence, the number of random variables drawn to determine how many reaction firings occurred in a leap is independent of the number of reaction firings.
Thus we effectively ``leap" over all of the reactions within a leap in few computational steps.
If molecular counts do not change by much within a leap then the fixed propensities are close to their correct SSA values and the approximation is good.

Our definition of a $\rho$-perturbation allows us to formally define  ``good".
We want to guarantee that the approximate process that tau-leaping actually simulates is a $\rho$-perturbation of the exact SSA process.
We can achieve this as follows.
If $\vec{x}$ is the state on which the leap started,
throughout the leap the simulated reaction propensities are fixed at their SSA propensities on $\vec{x}$: 
$a_j(\vec{x})$.
Then for any state $\vec{y}$ within the leap we want the correct SSA propensities $a_j(\vec{y})$ to satisfy the following \emph{$\rho$-perturbation constraint} ($0 < \rho < 1$):
$(1-\rho) a_j(\vec{y}) \leq a_j(\vec{x}) \leq (1+\rho) a_j(\vec{y})$.
As soon as we reach a state $\vec{y}$ for which this constraint is violated,
we start a new leap at $\vec{y}$ which will use simulated reaction propensities fixed at $a_j(\vec{y})$.
This ensures that at any time in the simulation, there is some $(1-\rho) \leq \xi_j(t) \leq (1+\rho)$ such that multiplying the correct SSA propensity of reaction $R_j$ by $\xi_j(t)$ yields the propensity of $R_j$ that the simulation algorithm is actually using.
Therefore, we actually simulate a $\rho$-perturbation,
and for $(\rho, \delta)$-robust SSA processes, 
the algorithm can be used to provably solve the prediction problem with error $\delta$.

Can we implement this simulation quickly, and, as promised, do little computation per leap?
Note that in order to limit the maximum propensity deviation in a leap, 
we need to make the leap duration be a random variable dependent upon the stochastic events in the leap.
If we evaluate $a_j(\vec{y})$ after each reaction occurrence in a leap to verify the satisfaction of the $\rho$-perturbation constraint, 
we do not save time over SSA.
However, we can avoid this by using a stricter constraint we call the \emph{$\{ \eps_{ij} \}$-perturbation constraint} ($0 < \eps_{ij} < 1$), defined as follows.
If the leap starts in state $\vec{x}$, reaction $R_j$ is allowed to change the molecular count of species $S_i$ by at most plus or minus $\eps_{ij} x_i$ within a leap.
Again, as soon as we reach a state $\vec{y}$ where this constraint is violated, we start a new leap at $\vec{y}$.%
\footnote{An added benefit of providing $\{\eps_{ij}\}$ bounds rather than $\rho$ as a parameter to the BTL algorithm is that it allows flexibility on the part of the user to assign less responsibility for a violation to a reaction that is expected to be fast compared to a reaction that is expected to be slow.
This may potentially speed up the simulation, while still preserving the $\rho$-perturbation constraint.}

For any $\rho$, we can find a set of $\{\eps_{ij}\}$ bounds such that satisfying the $\{ \eps_{ij} \}$-perturbation constraint satisfies the $\rho$-perturbation constraint.
In Appendix~\ref{sec:epsrho} we show that for any SCRN, 
the $\rho$-perturbation constraint is satisfied if $\eps_{ij} \leq \frac{3}{4M}(1-\sqrt{\frac{1+\rho/9}{1+\rho}})$, where $M$ is the number of reactions in the SCRN.

\begin{figure}[t]
\fbox{
\begin{minipage}{5.8in}
\begin{enumerate}
\item[0.] Initialize with time $t = t_0$ and the system's state $\vec{x} = \vec{x_0}$.
\item[1.] With the system in state $\vec{x}$ at time $t$, 
evaluate all the propensities $a_j$,
and determine firing bounds $b_j$ for all possible reactions,
where $b_j$ is the smallest positive integer such that $\abs{b_j \nu_{ij}} > \eps_{ij} x_i$ for some $S_i$.
\item[2.] Generate violating times $\tau_j  \sim \mathrm{Gamma} (b_j,a_j)$ for all possible reactions.
\item[3.] Find the first-violating reaction and set the step size to the time of the first violation: 
let $j^* = \argmin_j \{ \tau_j \}$ and $\tau = \tau_{j^*}$.
\item[4.] Determine the number of times each possible reaction occurred in interval $\tau$: for $j \neq j^*$, $n_j \sim \mathrm{Binomial} (b_j - 1, \tau/\tau_j)$; for $j^*$, $n_{j^*} = b_{j^*}$.
\item[5.] Effect the leap by replacing $t \leftarrow t + \tau$ and 
$\vec{x} \leftarrow \vec{x} + \sum_j \vec{\nu_j} n_j$.
\item[6.] Record $(\vec{x}, t)$ as desired.  Return to Step 1, or else end the simulation.
\end{enumerate}
\end{minipage}
}
\caption{The bounded tau-leaping (BTL) algorithm.   The algorithm is given the SCRN, the initial state $\vec{x_0}$, the volume $V$, and a set of perturbation bounds $\{ \eps_{ij} \} > 0$.
If the state at a specific time $t_f$ is desired, 
the algorithm checks if $t + \tau > t_f$ in step (3), and if so uses $\tau = t_f - \tau$, and treats all reactions as not first-violating in step (4).
Gamma$(n, \lambda)$ is a gamma distribution with shape parameter $n$ and rate parameter $\lambda$.  Binomial($n, p$) is a binomial distribution with number of trials $n$ and success probability $p$. 
}
\label{fig:alg}
\end{figure}

Simulating a leap such that it satisfies the $\{ \eps_{ij} \}$-perturbation constraint is easy and only requires drawing $M$ gamma and $M - 1$ binomial random variables.
Suppose the leap starts in state $\vec{x}$.
For each reaction $R_j$, let $b_j$ be the number of times $R_j$ needs to fire to cause a violation of the $\{\eps_{ij}\}$ bounds for some species.
Thus $b_j$ is the smallest positive integer such that $\abs{b_j \nu_{ij}} > \eps_{ij} x_i$ for some $S_i$.
To determine $\tau$, the duration of the leap, we do the following.
First we determine when each reaction $R_j$ would occur $b_j$ times, by drawing from a gamma distribution with shape parameter $b_j$ and rate parameter $a_j$.
This generates a time $\tau_j$ for each reaction.
The leap ends as soon as some reaction $R_j$ occurs $b_j$ times;
thus to determine the duration of the leap $\tau$ we take the minimum of the $\tau_j$'s.
At this point we know that the first-violating reaction $R_{j^*}$ --- the one with the minimum $\tau_{j^*}$ --- occurred $b_{j^*}$ times.  
But we also need to know how many times the other reactions occur.
Consider any other reaction $R_j$ $(j \neq j^*)$. 
Given that the $b_j$th occurrence of reaction $R_j$ would have happened at time $\tau_j$ had the leap not ended,
we need to distribute the other $b_j - 1$ occurrences to determine how many happen before time $\tau$.
The number of occurrences at time $\tau$ is given by the binomial distribution with number of trials $b_j(\vec{x}) - 1$ and success probability $\tau/\tau_j$.
This enables us to define BTL as shown in Figure~\ref{fig:alg}.

The algorithm is called ``bounded" tau-leaping because the deviations of reaction propensities within a leap are always bounded according to $\rho$.
This is in contrast with other tau-leaping algorithms, such as Gillespie's~\citep{gillespie06},
in which the deviations in reaction propensities are small with high probability, but not always, 
and in fact can get arbitrarily high if the simulation is long enough.
This allows BTL to satisfy our definition of a $\rho$-perturbation,
and permits easier analysis of the behavior of the algorithm (see next section).

As any algorithm exactly simulating a $\rho$-perturbation would, BTL naturally avoids negative concentrations.
Negative counts can occur only if an impossible reaction happens ---
in some state $\vec{x}$ reaction $R_j$ fires for which $a_j(\vec{x}) = 0$.
But since in a $\rho$-perturbation propensity deviations are multiplicative, in state $\vec{x}$, $a_j'(\vec{x}, t) = \xi_j(t) a_j(\vec{x}) = 0$ and so $R_j$ cannot occur.
Further, no matter how small the $\{\eps_{ij}\}$ bounds are, 
there is always at least one reaction per leap and thus BTL cannot take more steps than SSA.

On the negative side, in certain cases the BTL algorithm can take many more leaps than Gillespie's tau-leaping~\citep{Gillespie01,gillespie2003ils,gillespie06} and other versions.
Consider the case where there are two fast reactions that partially undo each others' effect (for example the reactions may be reverses of each other).
While both reactions may be occurring very rapidly, their propensities may be very similar (e.g.,~\citet{rathinam2007rem}).
Gillespie's tau-leaping will attempt to leap to a point where the molecular counts have changed enough according to the \emph{averaged} behavior of these reactions.
However, our algorithm considers each reaction separately and leaps to the point where the first reaction violates the bound on the change in a species in the absence of the other reactions.
Thus in this situation our algorithm would perform unnecessarily many leaps for the desired level of accuracy.

\subsection{Upper Bound on the Number of Leaps}   \label{sec:boundonleaps}

Suppose we fix some SCRN of interest, 
and run BTL on different initial states, 
volumes, and lengths of simulated time.
How does varying these parameters change the number of leaps taken by BTL?
In this section, we prove that no matter what the SCRN is, 
we can upper bound the number of leaps as a function of the total simulated time $t$, 
the volume $V$,
and the maximum total molecular count $m$ encountered during the simulation.
For simplicity we assume that all the $\eps_{ij}$ are equal to some global $\eps$.
(Alternatively, the theorem and proof can be easily changed to use min/max $\{\eps_{ij}\}$ values where appropriate.)

\begin{theorem}  \label{thm:algspeed}
For any SCRN $\mathcal{S}$ with $M$ species, 
any $\eps$ such that $0 < \eps < 1/(12 M)$, 
and any $\delta > 0$,
there are constants $c_1,c_2, c_3 > 0$ such that 
for any bounds on time $t$ and total molecular count $m$,
for any volume $V$ and any starting state,
after $c_1 \log{m} + c_2 \, t \, (C+c_3)$ leaps
where $C = m/V$,
either the bound on time or the bound on total molecular count will be exceeded
with probability at least $1-\delta$.
\end{theorem}
\begin{proof}
The proof is presented in Appendix~\ref{sec:leapbound}.
\end{proof}
Note that the upper bound on $\eps$ implies that the algorithm is exactly simulating some $\rho$-perturbation (see previous section).

Intuitively, a key idea in the proof of the theorem is that the propensity of a reaction decreasing a particular species is linear to the amount of that species (since the species must appear as a reactant).
This allows us to bound the decrease of any species if a leap is short.
Actually this implies that a short leap probably increases the amount of some species by a lot (some species must cause a violation --- if not by a decrease it must be by an increase).
This allows us to argue that if we have a lot of long leaps we exceed our time bound $t$ and if we have a lot of short leaps we exceed our bound on total molecular count $m$.
In fact because the effect of leaps is multiplicative, logarithmically many short leaps are enough to exceed $m$.

It is informative to compare this result with exact SSA, which in the worst case takes $O(1) \, m \, t \, (C+O(1))$ steps, since each reaction occurrence corresponds to an SSA step and
the maximum reaction propensity is $k_j m^2/V$ or $k_j m$.
Since $m$ can be very large, the speed improvement can be profound.

We believe, although it remains to be proven, that other versions of tau-leaping (see e.g.,~\citet{gillespie07} for a review) achieve the same asymptotic worst case number of leaps as our algorithm.

How much computation is required per each leap?
Each leap involves arithmetic operations on the molecular counts of the species, as well as drawing from a gamma and binomial distributions.
Since there are fast algorithms for obtaining instances of gamma and binomial random variables (e.g.,~\citet{ahrens1978ggv,kachitvichyanukul1988brv}), 
we do not expect a leap of BTL to require much more computation than other forms of tau-leaping, and should not be a major contributor to the total running time.
Precise bounds are dependent on the model of computation.
(In the next section we state reasonable asymptotic bounds on the computation time per leap for a randomized Turing machine implementation of BTL.)

\section{On the Computational Complexity of the Prediction Problem for Robust SSA Processes}    \label{sec:compcomplexity}

What is the computational complexity inherent in the prediction problem for robust SSA processes?
Is simulation with BTL a good procedure for solving the problem, or are there faster methods not involving simulation that somehow directly compute desired end-time probabilities without evaluating the entire trajectory?
We have shown that the number of leaps that BTL takes scales at most linearly with $t$ and $C$.
However, for some systems there are analytic shortcuts to determining the probability of being in $\Gamma$ at time $t$. 
For instance the ``exponential decay" SCRN consisting of the single reaction $S_1 \rightarrow S_2$ is easily solvable analytically~\citep{malek75}.
The calculation of the probability of being in any given state at any given time $t$ (among other questions) can be solved in time that grows minimally with $t$ and $C$.
Indeed, it may seem possible that robust SSA processes are somehow behaviorally weak and that their behavior can be easily predicted.

In order to be able to consider these questions formally, 
we specify our model of computation as being randomized Turing machines (see below).
We say that computation time polylogarithmic in the maximum total molecular count $m$ is \emph{efficient in $m$} 
(in fact $\log{m}$ computation time is required to simply read in the initial state of the SSA process and target state of the prediction problem).
In this section we prove that for any algorithm efficient in $m$ solving prediction problems for robust SSA processes,
there are prediction problems about such processes that cannot be solved faster than linear in $t$ and $C$.
We prove this result assuming a reasonable conjecture in computational complexity theory.
Then, 
with certain caveats regarding implementing BTL on a Turing machine, 
simulation with BTL is optimal in $t$ and $C$ for solving prediction problems for robust SSA processes
among algorithms efficient in $m$.

We use the standard model of computation which captures stochastic behavior: randomized Turing machines (TM).
A randomized TM is a non-deterministic TM\footnote{Arbitrary finite number of states and tapes.  Without loss of generality, we can assume a binary alphabet.} allowing multiple possible transitions at a point in a computation.  
The actual transition taken is uniform over the choices.
(See for example~\citet{sipser97} for equivalent formalizations.)
We say a given TM on a given input runs in computational time $t_{tm}$ if there is no set of random choices that makes the machine run longer.

We want to show that for some SCRNs, 
there is no method of solving the prediction problem fast, 
no matter how clever we are.
We also want these stochastic processes to be robust despite having difficult prediction problems.
We use the following two ideas.
First, a method based on~\citet{angluin06}
shows that predicting the output of given randomized TMs can be done by solving a prediction problem for certain robust SSA processes.
Second, an open conjecture, but one that is strongly compatible with the basic beliefs of computational complexity theory, 
bounds how quickly the output of randomized TMs can be determined.

Computational complexity theory concerns measuring how the computational resources required to solve a given problem scale with input size $n$ (in bits).
The two most prevalent efficiency measures are time and space 
--- the number of TM steps and the length of the TM tape required to perform the computation.
%\footnote{If $s(n)$ or $t(n)^\alpha$ is sublinear then the TM does not have enough time to read the input or space to hold the input.  
%If $t(n)$ is at least linear but $t(n)^\alpha$ is sublinear then the conjecture is obviously true. 
%However, it will not be useful to us since the new algorithm we'll make will have to use at least linear time and space to process the input.XXX}
%
We say a Boolean function $f(x)$ is \emph{probabilistically computable} by a TM $M$ in time $t(n)$ (where $n=\abs{x}$) and space $s(n)$ if $M(x)$ runs in time $t(n)$ using space at most $s(n)$, and with probability at least $2/3$ outputs $f(x)$.\footnote{Any other constant probability bounded away from $1/2$ will do just as well:
to achieve a larger constant probability of being correct, 
we can repeat the computation a constant number of times and take majority vote.}
A basic tenet of computational complexity is that allowing asymptotically more computation time $t(n)$ always expands the set of problems that can be solved.
Thus it is widely believed that for any (reasonable) $t(n)$, there are ``$t(n)$-hard" functions that can be probabilistically computed in $t(n)$ time, 
but not in asymptotically smaller time.\footnote{If we do not allow any chance of error, 
the corresponding statement is proven as the (deterministic) time hierarchy theorem~\citep{sipser97}.
Also see \citet{barak02,fortnow06} for progress in proving the probabilistic version.}
For our argument we will need such a $t(n)$-hard function, but one that does not require too much space.
Formally we assume the following \emph{hierarchy conjecture}:
\begin{conjecture}[(Probabilistic, Space-Limited) Time Hierarchy]     \label{con:hierarchyconjecture}
For any $\alpha < 1$, and polynomials $t(n)$ and $s(n)$ such that $t(n)^\alpha$ and $s(n)$ are at least linear,
there are Boolean functions that can be probabilistically computed within time and space bounds bounds $t(n)$ and $s(n)$, but not in time $O(1) t(n)^\alpha$ (with unrestricted space usage).
\end{conjecture}

Intuitively, we take a Boolean function that requires $t(n)$ time and embed it in a chemical system in such a way that solving the prediction problem is equivalent to probabilistically computing the function. 
The conjecture implies that we cannot solve the prediction problem fast enough to allow us to solve the computational problem faster than $t(n)$.
Further, since the resulting SSA process is robust, the result lower-bounds the computational complexity of the prediction problem for robust processes.
Note that we need a time hierarchy conjecture that restricts the space usage and talks about probabilistic computation because it is impossible to embed a TM computation in an SCRN such that its computation is error free~\citep{shukipaper}, and further such embedding seems to require more time as the space usage increases.

The following theorem lower-bounds the computational complexity of the prediction problem.
The bound holds even if we restrict ourselves to robust processes.
It shows that this computational complexity is at least linear in $t$ and $C$,
as long as the dependence on $m$ is at most polylogarithmic.
It leaves the possibility that there are algorithms for solving the prediction problem that require computation time more than polylogarithmic in $m$ but less than linear in $t$ or $C$.
Let the prediction problem be specified by giving the SSA process (via the initial state and volume),
the target time $t$, and
the target outcome $\Gamma$ in some standard encoding such that whether a state belongs to $\Gamma$ can be computed in time polylogarithmic in $m$.
\begin{theorem}   \label{thm:optimality}
Fix any perturbation bound $\rho > 0$ and $\delta > 0$.
Assuming the hierarchy conjecture (Conjecture~\ref{con:hierarchyconjecture}),
there is an SCRN $\mathcal{S}$ such that
for any prediction algorithm $\mathcal{A}$ and constants $c_1, c_2, \beta, \eta, \gamma >0$,
there is an SSA process $\mathcal{C}$ of $\mathcal{S}$ and a 
$(m, t, C, 1/3)$-prediction problem $\mathcal{P}$ of $\mathcal{C}$ 
such that $\mathcal{A}$ cannot solve $\mathcal{P}$ in computational time 
$c_1\, (\log m)^{\beta} \, t^\eta \, (C+c_2)^\gamma$ if $\eta < 1$ or  $\gamma < 1$.
Further, $\mathcal{C}$ is $(\rho, \delta)$-robust with respect to $\mathcal{P}$.
\end{theorem}
\begin{proof}
The proof is presented in Appendix~\ref{sec:optimality}.
\end{proof}

%In the previous section, we have derived an upper bound on the number of leaps that our algorithm takes.
%However, we need to address how the idealized bounded-tau leaping algorithm presented in Section~\ref{sec:algorithm} can be implemented on a randomized TM which allows only finite precision arithmetic and a restricted model of randomness generation.
%We have to deal with round-off error and approximate gamma and binomial random number generators, whose effect on the probability of outcome is difficult to track formally.
%Further, the computational complexity of these operations is a function of the bits of precision and is complicated to rigorously bound.

In Appendix~\ref{sec:TMimplementation}, we argue that
BTL on a randomized TM runs in total computation time 
$$O(1)((\log(m))^{O(1)} +l) \, t \, (C+ O(1))$$
where, in each leap, polylogarithmic time in $m$ is required for arithmetic manipulation of molecular counts,
and $l$ captures the extra computation time required for the real number operations and drawing from the gamma and binomial distributions.
Here $l$ is potentially a function of $m$, $V$, $t$, and the bits of precision used,
but assuming efficient methods for drawing the random variables
$l$ is likely very small compared to the total number of leaps.
Then, assuming we can ignore errors introduced due to finite precision arithmetic and approximate random number generation,
simulation with BTL is an asymptotically optimal way in $t$ and $C$ of solving the prediction problem for robust processes among methods efficient in $m$.

\section{Discussion}

The behavior of many stochastic chemical reaction networks does not depend crucially on getting the reaction propensities exactly right,
prompting our definition of $\rho$-perturbations and $(\rho, \delta)$-robustness.
A $\rho$-perturbation of an SSA process is a stochastic process with stochastic deviations of the reaction propensities from their correct SSA values.
These deviations are multiplicative and bounded between $1-\rho$ and $1+\rho$.
If we are concerned with how likely it is that the SSA process is in a given state at a given time,
then $(\rho, \delta)$-robustness captures how far these probabilities may deviate for a $\rho$-perturbation.

We formally showed that predicting the behavior of robust processes does not require simulation of every reaction event.
Specifically, we described a new approximate simulation algorithm called bounded tau-leaping (BTL) that simulates a certain $\rho$-perturbation as opposed to the exact SSA process.
The accuracy of the algorithm in making predictions about the state of the system at given times is guaranteed for $(\rho,\delta)$-robust processes.
We proved an upper bound on the number of leaps of BTL that helps explain the savings over SSA.
The bound is a function of the desired length of simulated time $t$, volume $V$, and maximum molecular count encountered $m$.
This bound scales linearly with $t$ and $C=m/V$, but only logarithmically with $m$,
while the total number of reactions (and therefore SSA steps) may scale linearly with $t$, $C$, and $m$.
When total concentration is limited, but the total molecular count is large, 
this represents a profound improvement over SSA.
Because the number of BTL leaps scales only logarithmically with $m$,
BTL asymptotically nears the speed of mass action ODE solvers --- which have no dependence on $m$.
We also argue that asymptotically as a function of $t$ and $C$ our algorithm is optimal in as far as no algorithm can achieve sublinear dependence of the number of leaps on $t$ or $C$.
This result is proven based on a reasonable assumption in computational complexity theory.
Unlike Gillespie's tau-leaping~\citep{gillespie06}, 
our algorithm seems better suited to theoretical analysis.
Thus while we believe other versions of tau-leaping have similar worst-case running times, 
the results analogous to those we obtain for BTL remain to be proved.

Our results can also be seen to address the following question.
If concerned solely with a particular outcome rather than with the entire process trajectory, 
can one always find certain shortcuts to determine the probability of the outcome
without performing a full simulation?
Since our lower bound on computation time scales linearly with $t$, 
it could be interpreted to mean that,
except in problem-specific cases, 
there is no shorter route to predicting the outcomes of stochastic chemical processes than via simulation.
This negative result holds even restricting to the class of robust SSA processes.

While the notion of robustness is a useful theoretical construct,
how practical is our definition in deciding whether a given system is suitable to approximate simulation via BTL or not?
We prove that for the class of monotonic SSA processes,
robustness is guaranteed at all times when in the SSA process the outcome probability is stable over an interval of time determined by $\rho$.
However, it is not clear how this stability can be determined without SSA simulation.
Even worse, few systems of interest are monotonic.
Consequently, it is compelling to develop techniques to establish robustness for more general classes of systems.
A related question is whether it is possible to connect our notion of robustness to previously studied notions in mass action stability analysis~\citep{horn1972gma,sontag2007man}.

%What if we had chosen a different definition of robustness?  (Say with respect to some more complex function of trajectory -- stronger kind of robustness.)

\appendix

\section{Appendix}

\subsection{Enforcing the $\rho$-Perturbation Constraint by the $\{\eps_{ij}\}$-Perturbation Constraint}  
\label{sec:epsrho}

Recall that in Section~\ref{sec:algorithm} we introduced two constraints bounding the number of reaction events within a leap.
If $\vec{x}$ is the state at the beginning of the leap,
the $\rho$-perturbation constraint is satisfied if for every reaction $R_j$,
$(1-\rho) a_j(\vec{y}) \leq a_j(\vec{x}) \leq (1+\rho) a_j(\vec{y})$ for any state $\vec{y}$ within the leap.
The $\{\eps_{ij}\}$-perturbation constraint is satisfied if 
no reaction $R_j$ changes the molecular count of species $S_i$ by more than plus or minus $\eps_{ij} x_i$ within the leap.
For a given $\rho$, 
we would like to find an appropriate $\{\eps_{ij}\}$-perturbation constraint to use in the BTL algorithm
such that we satisfy the $\rho$-perturbation constraint, thereby ensuring that we are exactly simulating some $\rho$-perturbation.
To avoid making the $\{\eps_{ij}\}$-perturbation constraint tighter than necessary requires knowledge of the exact reactions in the given SCRN.
Nevertheless, worst-case analysis below shows that setting $\eps_{ij} \leq \frac{3}{4M}(1-\sqrt{\frac{1+\rho/9}{1+\rho}})$ works for any SCRN of $M$ reactions.

If $\eps_{ij} = \eps$
then, for any SCRN with $M$ reactions, 
the maximum change of any species $S_i$ within a leap allowed by the $\{\eps_{ij}\}$-perturbation constraint is plus or minus $M \eps x_i$.
We want to find an $\eps > 0$ such that if the changes to all species stay within the $M \eps$ bounds, 
then no reaction violates the $\rho$-perturbation constraint.
Consider a bimolecular reaction $R_j$: $2 S_i \rightarrow \dots$ first.
The algorithm simulates its propensity as $a_j(\vec{x}) = k_j x_i (x_i-1) / V$ throughout the leap.
If $x_i = 0$ or $1$, then $a_j(\vec{x}) = 0$, and as long as $M \eps < 1$, then still $a_j(\vec{y}) = 0$, satisfying the $\rho$-perturbation constraint for $R_j$.
Otherwise, if $x_i \geq 2$,
then the SSA propensity at state $\vec{y}$ within the leap is 
$a_j(\vec{y}) = k_j y_i (y_i - 1)/V \leq k_j (1+ M \eps) x_i ((1+ M \eps) x_{i} - 1) / V$,
and so the left half of the $\rho$-perturbation constraint 
$(1-\rho) a_j(\vec{y}) \leq a_j (\vec{x})$ 
is satisfied 
if $(1-\rho)(1+ M \eps) x_i ((1+ M \eps) x_{i} - 1) \leq  x_i (x_i - 1)$.
Similarly, 
%$a_j(\vec{y}) = k_j y_i (y_i - 1)/V \geq k_j (1- M \eps) x_i ((1- M \eps) x_{i} - 1) / V$
%and 
the right half of the $\rho$-perturbation constraint $a_j (\vec{x})  \leq (1+\rho) a_j(\vec{y})$ 
is satisfied 
if $(1+\rho)(1- M \eps) x_i ((1- M \eps) x_{i} - 1) \geq  x_i (x_i - 1)$.
These inequalities are satisfied for $x_i \geq 2$ when
$\eps \leq \frac{3}{4M}(1-\sqrt{\frac{1+\rho/9}{1+\rho}})$ 
(which also ensures that $M \eps < 1$).

%This setting of $\eps$ also works for other reaction types.
In a likewise manner, for a unimolecular reaction $R_j$: $S_i \rightarrow \dots$,
the $\rho$-perturbation constraint is satisfied if 
$(1-\rho)(1+ M \eps) x_i \leq  x_i$
and
$(1+\rho)(1- M \eps) x_i \geq  x_i$,
and for a bimolecular reaction $R_j$: $S_i + S_{i'} \rightarrow \dots$,
the constraint is satisfied if
$(1-\rho)(1+ M \eps)^2 x_i x_{i'} \leq  x_i x_{i'}$ and
$(1+\rho)(1- M \eps)^2 x_i x_{i'} \geq  x_i x_{i'}$.
It is easy to see that setting $\eps$ as above also satisfies the inequalities for these reaction types.
%Thus for any $\rho$ we know how small $\eps$ needs to be such that 
%satisfying the corresponding $\{ \eps_{ij} \}$-perturbation constraint 
%ensures that we are exactly simulating some $\rho$-perturbation.

Throughout the paper we assume that $\rho$, $\eps$ or $\{ \eps_{ij} \}$ are fixed and most of our asymptotic results do not show dependence on these parameters.
Nonetheless, we can show that for a fixed SCRN and for small enough $\rho$, $\eps$ can be within the range $O(1)\rho \leq \eps \leq O(1) \rho$ and thus scales linearly with $\rho$.
Therefore, in asymptotic results, the dependence on $\eps$ and $\rho$ can be interchanged.
Specifically, the $\eps$ dependence explored in Appendix~\ref{sec:leapbound} can be equally well expressed as a dependence on $\rho$.

\subsection{Proof of Theorem~\ref{thm:algspeed}: Upper Bound on the Number of Leaps}   \label{sec:leapbound}

In this section we prove Theorem~\ref{thm:algspeed} from the text, 
which upper-bounds the number of leaps BTL takes
as a function of $m$, $t$, and $C$:

\begin{theorem*}
For any SCRN $\mathcal{S}$ with $M$ species, 
any $\eps$ such that $0 < \eps < 1/(12 M)$, 
and any $\delta > 0$,
there are constants $c_1,c_2, c_3 > 0$ such that 
for any bounds on time $t$ and total molecular count $m$,
for any volume $V$ and any starting state,
after $c_1 \log{m} + c_2 \, t \, (C+c_3)$ leaps
where $C = m/V$,
either the bound on time or the bound on total molecular count will be exceeded
with probability at least $1-\delta$.
\end{theorem*}

We prove a more detailed bound than stated in the theorem above which explicitly shows the dependence 
on $\eps$ hidden in the constants.
Also since we introduce the asymptotic results only the end of the argument, 
the interested reader may easily investigate the dependence of the constants on other parameters of the SCRN such as $N$, $M$, $\nu_{ij}$, and $k_j$.
We also show an approach to probability $1$ that occurs exponentially fast as the bound increases:
if the bound above evaluates to $n$, then the probability that the algorithm does not exceed $m$ or $t$ in $n$ leaps is at most $2 e^{-O(1) n}$.

Our argument starts with a couple of lemmas.
Looking within a single leap, the first lemma bounds the decrease in the molecular count of a species due to a given reaction as a function of time.
The argument is essentially that for a reaction to decrease the molecular count of a species, 
that species must be a reactant, and therefore the propensity of the reaction is proportional to its molecular count.
Thus we see a similarity to an exponential decay process and use this to bound the decrease.
Note that a similar result does not hold for the \emph{increase} in the molecular count of a species,
since the molecular count of the increasing species need not be in the propensity function.\footnote{If a reaction is converting a populous species to a rare one, 
the rate of the increase of the rare species can be proportional to $m$ times its molecular count.
The rate of decrease, however, is always proportional to the molecular count of the decreasing species, or proportional to $C$ times the molecular count of the decreasing species (as we'll see below).}
Then the second lemma uses the upper bound on how fast a species can decrease (the first lemma), 
together with the fact that in a leap some reaction must change some species by a relatively large amount, 
to classify leaps into those that either (1) take a long time or (2) increase some species significantly without decreasing any other species by much.
Finally we show that this implies that if there are too many leaps we either violate the time bound or the total molecular count bound.

For the following, values $f$ and $g$ will be free parameters to be determined later. 
It helps to think of them as $0 < f \ll g \ll 1$.
How long does it take for a reaction to decrease $x_i$ by $g$th fraction of the violation bound $\eps x_i$?
The number of occurrences of $R_j$ to decrease $x_i$ by $g \eps x_i$ or more is at least $g \eps x_i / \abs{\nu_{ij}}$.
The following lemma bounds the time required for these many occurrences to happen.
\begin{lemma}   \label{lem:keyprob}
Take any $f$ and $g$ $(0 < f, g < 1)$,
any reaction $R_j$ and species $S_i$ such that $\nu_{ij} < 0$,
any state $\vec{x}$,
and any $\eps$.
Assuming that the propensity of $R_j$ is fixed at $a_j(\vec{x})$,
with probability at least $1-f/g$,
fewer than $g \eps x_i/\abs{\nu_{ij}}$ occurrences of $R_j$ happen
in time $f \eps /(\abs{\nu_{ij}} k_j)$ if $R_j$ is unimolecular,
or time $f \eps /(\abs{\nu_{ij}} k_j C)$ if $R_j$ is bimolecular.
\end{lemma}
\begin{proof}
For reaction $R_j$ to decrease the amount of $S_i$, it must be that $S_i$ is a reactant, and thus $x_i$ is a factor in the propensity function.
Suppose $R_j$ is unimolecular.
Then $a_j = k_j x_i$ and the expected number of occurrences of $R_j$ in time $f  \frac{\eps}{\abs{\nu_{ij}} k_j}$ is $a_j  f  \frac{\eps}{\abs{\nu_{ij}} k_j}     \leq     f  \frac{\eps x_i}{\abs{\nu_{ij}}}$.
The desired result then follows from Markov's inequality.
If $R_j$ is bimolecular with $S_i \neq S_{i'}$ being the other reactant
then $a_j = k_j \frac{x_i x_{i'}}{V}$; alternatively, $a_j = k_j \frac{x_i (x_i-1)}{V}$ if $R_j$ is bimolecular with identical reactants.
In general for bimolecular reactions $a_j \leq  k_j x_i C$.
So the expected number of occurrences of $R_j$ in time $f  \frac{\eps}{\abs{\nu_{ij}} k_j C}$ is $a_j  f  \frac{\eps}{\abs{\nu_{ij}} k_j C}     \leq     f  \frac{\eps x_i}{\abs{\nu_{ij}}}$.
The desired result follows as before.
\end{proof}

Let time $\tilde\tau$ be the minimum over all reactions $R_j$ and $S_i$ such that $\nu_{ij} < 0$ of
$1/(\abs{\nu_{ij}} k_j)$ if $R_j$ is unimolecular,
or $1/(\abs{\nu_{ij}} k_j C)$ if $R_j$ is bimolecular.
We can think of $\tilde\tau$ setting the units of time for our argument.
The above lemma implies that with probability at least $1-f/g$ no reaction decreases $x_i$ by $g \eps x_i$ or more within time $f \eps \tilde\tau$.
The following lemma defines typical leaps; they are of two types: long or $S_i$-increasing.
Recall $M$ is the number of reaction channels and $N$ is the number of species.

\begin{lemma}    \label{lem:typical}
(Typical leaps).
For any $f$ and $g$ $(0 < f, g < 1)$,
and for any $\eps$, 
with probability at least $1-N M f/g$
one of the following is true of a leap:
\begin{itemize}
\item[1.](long leap) $\tau > f \eps \tilde{\tau}$
\item[2.]($S_i$-increasing leap) $\tau \leq f \eps \tilde{\tau}$, and the leap increases some species $S_i$ at least as $x_i \mapsto x_i + \lceil \eps x_i \rceil - g M \eps x_i$,
while no species $S_{i'}$ decreases as much as $x_{i'} \mapsto x_{i'} - g M \eps x_{i'}$.
\end{itemize}
\end{lemma}
\begin{proof}
By the union bound over the $M$ reaction channels and the $N$ species, 
Lemma~\ref{lem:keyprob} implies that the probability that \emph{some} reaction decreases the amount of \emph{some} species $S_i$ by $g \eps x_i$ or more in time $f \eps \tilde\tau$ is at most $N M f/g$.
Now suppose this unlucky event does not happen.
Then if the leap time is $\tau \leq f \eps \tilde\tau$, 
no decrease is enough to cause a violation of the deviation bounds,
and thus it must be that some reaction $R_{j}$ increases some species $S_{i}$ by more than $\eps x_{i}$.
(Since $R_j$ must occur an integer number of times,
it actually must increase $S_i$ by $\lceil \eps x_{i} \rceil$ or more.)
Since no reaction decreases $S_i$ by $g \eps x_i$ or more, we can be sure that $S_i$ increases at least by $\lceil \eps x_i \rceil - g M \eps x_i$.
\end{proof}

\begin{lemma}   \label{lem:worstdecrease}
For any species $S_i$, a leap decreases $S_i$ at most as $x_i \mapsto x_i - M \lfloor \eps x_i \rfloor - 2$.
\end{lemma}
\begin{proof}
At most $M$ reactions may be decreasing $S_i$.
A reaction can decrease $S_i$ by as much as $\lfloor \eps x_i \rfloor$ without causing a violation of the deviation bounds.
The last reaction firing that causes the violation of the deviation bounds ending the leap uses up at most $2$ molecules of $S_i$ (since reactions are at most bimolecular).
\end{proof}

Note that a similar lemma does not hold for Gillespie's tau-leaping algorithms~\citep{Gillespie01,gillespie2003ils,gillespie06}
because the number of reaction firings in a leap can be only bounded probabilistically.
With some small probability a leap can result in ``catastrophic" changes to some molecular counts.
Since with enough time such events are certain to occur,
the asymptotic analysis must consider them.
Consequently, asymptotic results analogous to those we derive in this section remain to be proved for tau-leaping algorithms other than BTL.

Our goal now is to use the above two lemmas to argue that if we have a lot of leaps, 
we would either violate the molecular count bound (due to many $S_i$-increasing leaps for the same $S_i$), or violate the time bound (due to long leaps).
Let $n$ be the total number of leaps.
By Hoeffding's inequality, with probability at least $1-2 e^{-2 n (NM f/g)^2}$ (i.e.,\ exponentially approaching $1$ with $n$), the total number of atypical steps is bounded as:
\begin{equation}
\mbox{[\# of atypical leaps]} < 2 n NM f/g.
\label{eqn:atypicalsteps}
\end{equation}
Further, in order not to violate the time bound $t$, the number of long steps can be bounded as:
\begin{equation}
\mbox{[\# of long leaps]} \leq t / (f \eps \tilde\tau).
\label{eqn:longsteps}
\end{equation}

How can we bound the number of the other leaps ($S_i$-increasing, for some species $S_i$)?
Our argument will be that having too many of such leaps results in 
an excessive increase of a certain species,
thus violating the bound on the total molecular count.
We start by choosing an $S_i$ for which there is the largest number of $S_i$-increasing steps.
Since there are $N$ species, there must be a species $S_i$ for which
\begin{equation}
\mbox{[\# of $S_i$-increasing leaps]} > \frac{1}{N}\sum_{S_{i'} \neq S_{i}} \mbox{[\# of $S_{i'}$-increasing leaps]}.
\label{eqn:siincreasingstep}
\end{equation}

At this point, it helps to develop an alternative bit of notation labeling the different kinds of leaps with respect to the above-chosen species $S_i$ to indicate how much $x_i$ may change in the leap.
Since our goal will be to argue that the molecular count of $S_i$ must be large,
we would like to lower-bound the increase in $S_i$ and upper-bound the decrease.
An atypical leap or a long leap we label ``$\bigdownstep$".
By Lemma~\ref{lem:worstdecrease} these leaps decrease $S_i$ \emph{at most} as $x_i \mapsto x_i - M \lfloor \eps x_i \rfloor - 2$.
An $S_i$-increasing leap we label ``$\upstep$".
Finally, an $S_{i'}$-increasing leap for $S_{i'} \neq S_i$ we label ``$\smalldownstep$".
By Lemma~\ref{lem:typical}, $\upstep$ leaps increase $S_i$ \emph{at least} as $x_i \mapsto x_i + \lceil \eps x_i \rceil - g M \eps x_i$,
while $\smalldownstep$ leaps decrease $S_i$ \emph{by less} than $x_{i} \mapsto x_{i} - g M \eps x_{i}$.

We would like to express these operations purely in a multiplicative way so that they become commutative, allowing for bounding their total effect on $x_i$ independent of the order in which these leaps occurred but solely as a function of the number of each type.
Further, the multiplicative representation of the leap effects is important because we want to bound the number of leaps logarithmically in the maximum molecular count.
Note that $\bigdownstep$ leaps cause a problem because of the subtractive constant term, 
and $\upstep$ leaps cause a problem because if $x_i$ drops to $0$ 
multiplicative increases are futile.
Nonetheless, for the sake of argument suppose we knew that throughout the simulation $x_i \geq 3$.
Then assuming $\eps \leq 1/(12 M)$,
we can bound the largest decrease due to a $\bigdownstep$ leap multiplicatively as $x_i \mapsto (1/4) \, x_i$.
Further, we lower-bound the increase due to a $\upstep$ leap as $x_i \mapsto (1+(1-gM)\eps) x_i$.
Then the lower bound on the final molecular count of $S_i$ and therefore the total molecular count is
\begin{equation}
3(1+(1-gM)\eps)^{n^\upstep} (1-g M \eps)^{n^\smalldownstep} (1/4)^{n^\bigdownstep}  \leq m.
\label{eqn:firstmultiplicativebound}
\end{equation}
This implies an upper bound on the number of $\upstep$ leaps, 
that together with (eqns.~\ref{eqn:atypicalsteps})--(\ref{eqn:siincreasingstep}) 
provides an upper bound on the total number of leaps, 
as we'll see below.

However, $x_i$ might dip below $3$ (including at the start of the simulation).
We can adjust the effective number of $\upstep$ leaps to compensate for these dips.
We say a leap is in a dip if it starts at $x_i < 3$. 
Observe that the first leap in a dip starts at $x_i < 3$ while the leap after a dip starts at $x_i \geq 3$.
Thus, unless we end in a dip, cutting out the leaps in the dips can only decrease our lower bound on the final $x_i$. 
We'll make an even looser bound and modify~(\ref{eqn:firstmultiplicativebound}) simply by removing the contribution of the $\upstep$ leaps that are in dips.\footnote{We know we cannot end in a dip if the resulting bound evaluates to $3$ or more.
Thus technically we assume $m \geq 3$ for the bound to be always valid.}
How many $\upstep$ leaps can be in dips?
First let us ensure $g < 1/(3 M)$. 
Then since a $\smalldownstep$ leap decreases $x_i$ by less than $g M \eps x_i < x_i/3$,
and the decrease amount must be an integer,
a $\smalldownstep$ leap cannot bring $x_i$ below $3$ starting at $x_i \geq 3$.
Thus if we start at $x_i \geq 3$ a $\bigdownstep$ leap must occur before we dip below $3$. 
Thus the largest number of dips is $n^\bigdownstep + 1$ (adding $1$ since we may start the simulation below $3$).
Let $n_d^\upstep$ and $n_d^\bigdownstep$ be the number of $\upstep$ and $\bigdownstep$ leaps in the $d$th dip (we don't care about $\smalldownstep$ leaps in a dip since they must leave $x_i$ unchanged).
Then $n_d^\upstep < 2 n_d^\bigdownstep + 3$ and $\sum_d n_d^\upstep <  \sum_d 2 n_d^\bigdownstep + \sum_d 3  \leq 2 n^\bigdownstep + 3(n^\bigdownstep + 1) = 5 n^\bigdownstep + 3$.
Therefore, the adjusted bound~(\ref{eqn:firstmultiplicativebound}) becomes:
$3(1+(1-gM)\eps)^{n^\upstep - 5 n^\bigdownstep - 3} (1-g M \eps)^{n^\smalldownstep} (1/4)^{n^\bigdownstep}  \leq m$.
For simplicity, we use the weaker bound 
\begin{equation}
3(1+(1-gM)\eps)^{n^\upstep} (1-g M \eps)^{n^\smalldownstep} (1/4)^{6 n^\bigdownstep+3} \leq m.
\label{eqn:secondmultiplicativebound}
\end{equation}

In order to argue that this bounds the number of $\upstep$ leaps, 
we need to make sure the $\smalldownstep$ leaps and the $\bigdownstep$ leaps don't cancel out the effect of the $\upstep$ leaps.
By inequality~\ref{eqn:siincreasingstep} we know that $n^\smalldownstep < N n^\upstep$.
If we can choose $g$ to be a small enough constant such that more than $N$ $\smalldownstep$ leaps are required to cancel the effect of a $\upstep$ leap we would be certain the bound increases exponentially with $n^\upstep$ without caring about $\smalldownstep$ leaps.
Specifically, we choose a $g$ small enough such that $(1+(1-gM)\eps)(1-g M \eps)^N \geq 1+\eps/2$.
For example we can let $g = \frac{1}{M}(1-(9/10)^{1/N})$.\footnote{
Since $g \leq 1/(3 M)$, make the simplification $(1+(1-g M) \eps) \geq (1+2\eps/3)$ and solve for $g$.
The solution is minimized when $\eps = 1$.
}
Note that $g$ depends only on constants $N$ and $M$ and is independent of $\eps$.
The bound then becomes $3 (1+\eps/2)^{n^\upstep} (1/4)^{6 n^\bigdownstep + 3}$.

Thus finally we have the following system of inequalities that are satisfied with probability exponentially approaching $1$ as $n \rightarrow \infty$:
\begin{equation}
n = n^\upstep + n^\smalldownstep + n^\bigdownstep
\end{equation}
\begin{equation}
n^\bigdownstep \leq t/(f \eps \tilde\tau) + 2 n N M f/g
\end{equation}
\begin{equation}
n^\smalldownstep < N n^\upstep
\end{equation}
\begin{equation}
3 (1+\eps/2)^{n^\upstep} (1/4)^{6n^\bigdownstep + 3} \leq m.
\end{equation}

Solving for $n$ we obtain\footnote{Logarithms are base-2.} 
$$n \leq \frac{h \log(m/3) + (12 h + 1) t/(f \eps \tilde\tau) + 6h}       {(1-24 h NM f/g)}$$
if $(1-24hf/g) > 0$ where 
$h = (N+1)/\log(1+\eps/2)$
(also recall $g = \frac{1}{M}(1-(9/10)^{1/N})$).
To ensure this we let $f \leq g/(48hNM)$.
Then with probability exponentially approaching $1$ as $n$ increases,
$$n \leq 2\log(m/3) + 96(12h+1) t h/(g \eps \tilde\tau) + 12h.$$

Asymptotically as $\eps \rightarrow 0, m \rightarrow \infty, t \rightarrow \infty$ with the system of chemical equations being fixed,
we have $g = O(1)$, $h \leq O(1)/\eps$, and write the above as
$n \leq O(1) (1/\eps) \log{m} + O(1) (1/\eps)^3 t/\tilde\tau.$
Recall our unit of time $\tilde\tau$ was defined to be the minimum over all reactions $R_j$ and species $S_i$ such that $\nu_{ij} < 0$ of
$1/(\abs{\nu_{ij}} k_j)$ if $R_j$ is unimolecular,
or $1/(\abs{\nu_{ij}} k_j C)$ if $R_j$ is bimolecular.
No matter what $C$ is, we can say $\tilde\tau \geq 1/(O(1) C+O(1))$.
Thus we can write the above as 
$$n \leq O(1) (1/\eps) \log{m} + O(1) (1/\eps)^3 t (C+O(1)).$$
For any $\delta$, we can find appropriate constants such that the above bound is satisfied with probability at least $1-\delta$.

This bound on the number of leaps has been optimized for simplicity of proof rather than tightness.
A more sophisticated analysis can likely significantly decrease the numerical constants.
Further, we believe the cubic dependence on $1/\eps$ in the time term is excessive.\footnote{
The cubic dependence on $1/\eps$ in the time term is due to having to decrease the probability of an atypical step as $\eps$ decreases.
It may be possible to reduce the cubic dependence to a linear one by moving up the boundary between a dip and the multiplicative regime as a function of $\eps$ rather than fixing it at $3$.
The goal is to replace the constant base $(1/4)^{O(1)n^\bigdownstep+O(1)}$ term with a $(1-O(1) \eps)^{O(1)n^\bigdownstep+O(1)}$ term.
Then the effect of a $\bigdownstep$ leap would scale with $\eps$,  as does the effect of an $\upstep$ leap.}

\subsection{Proving Robustness by Monotonicity}       \label{sec:monotonicity}
In this section we develop a technique that can be used to prove the robustness of certain SSA processes.
We use these results to prove the robustness of the example in Section~\ref{sec:examples} as well as of the construction of \citet{angluin06} simulating a Turing machine in Appendix~\ref{sec:robustembedding}.

Since $\rho$-perturbations are not Markovian, it is difficult to think about them.
Can we use a property of the original SSA process that would allow us to prove robustness without referring to $\rho$-perturbations at all?

Some systems have the property that every reaction can only bring the system ``closer" to the outcome of interest (or at least ``no futher").  
Formally, we say an SSA process is \emph{monotonic} for outcome $\Gamma$ if for all reachable states $\vec{x}, \vec{y}$ such that there is a reaction taking $\vec{x}$ to $\vec{y}$,
and for all $t$,
the probability of reaching $\Gamma$ within time $t$ starting at $\vec{y}$ 
is at least the probability of reaching $\Gamma$ within time $t$ starting at $\vec{x}$.
Note that by this definition $\Gamma$ must be absorbing.
Intuitively, perturbation of propensities in monotonic systems only change how fast the system approaches the outcome.
Indeed, we can bound the deviations in the outcome probability of any $\rho$-perturbation at any time 
by two specific $\rho$-perturbations, which are the maximally slowed down and sped up versions of the original process.
This implies that monotonic SSA processes are robust at all times $t$ when the outcome probability does not change quickly with $t$, and thus slowing down or speeding up the SSA process does not significantly affect the probability of the outcome.

For an SSA process or $\rho$-perturbation $\mathcal{C}$ and set of states $\Gamma$, 
define $F^\Gamma(\mathcal{C},t)$ to be the probability of being in $\Gamma$ at time $t$. 
For SSA process $\mathcal{C}$, let $\tilde{\mathcal{C}}^{-\rho}$ be the $\rho$-perturbation defined by the constant deviations $\xi_j(t) = 1-\rho$.
Similarly, let $\tilde{\mathcal{C}}^{+\rho}$ be the $\rho$-perturbation defined by the constant deviations $\xi_j(t) = 1+\rho$.

\begin{lemma}  \label{lem:monotonicity}
If an SSA process $\mathcal{C}$ is monotonic for outcome $\Gamma$,
then for any $\rho$-perturbation $\tilde\mathcal{C}$ of $\mathcal{C}$, 
$F^\Gamma(\tilde{\mathcal{C}}^{-\rho}, t) \leq F^\Gamma(\tilde\mathcal{C},t) \leq F^\Gamma(\tilde\mathcal{C}^{+\rho}, t)$.
\end{lemma}
\begin{proof}
If an SSA process is monotonic,
allowing extra ``spontaneous" transitions (as long as they are legal according to the SSA process)
cannot induce a delay in entering $\Gamma$. 
We can decompose a perturbation with $\xi_j(t) \geq 1$ as the SSA process combined with some extra probability of reaction occurrence in the next interval $dt$.
Thus,
for a perturbation $\tilde\mathcal{C}$ of a monotonic SSA process $\mathcal{C}$ in which $\xi_j(t) \geq 1$,
we have $F^\Gamma(\mathcal{C}, t) \leq F^\Gamma(\tilde\mathcal{C},t)$.
By a similar argument, if $\tilde\mathcal{C}$ has $\xi_j(t)  \leq 1$,
then $F^\Gamma(\tilde\mathcal{C}, t) \leq F^\Gamma(\mathcal{C}, t)$.
Now $\tilde{\mathcal{C}}^{-\rho}$ and $\tilde{\mathcal{C}}^{+\rho}$ are themselves monotonic SSA processes ($\mathcal{C}$ scaled in time).
Then by the above bounds, for any $\rho$-perturbation $\tilde\mathcal{C}$ of $\mathcal{C}$
we have $F^\Gamma(\tilde{\mathcal{C}}^{-\rho}, t)  \leq F^\Gamma(\tilde\mathcal{C}, t) \leq F^\Gamma(\tilde{\mathcal{C}}^{+\rho}, t)$.
\end{proof}

Since $\tilde{\mathcal{C}}^{-\rho}$ and $\tilde{\mathcal{C}}^{+\rho}$ are simply the original SSA process $\mathcal{C}$ scaled in time by a factor of $1/(1+\rho)$ and $1/(1-\rho)$, respectively, 
we can write the bound of the above lemma as $F^\Gamma(\mathcal{C}, t/(1+\rho)) \leq F^\Gamma(\tilde\mathcal{C},t) \leq F^\Gamma(\mathcal{C}, t/(1-\rho))$.
Rephrasing Lemma~\ref{lem:monotonicity}:

\begin{corollary}
If an SSA process $\mathcal{C}$ is monotonic for outcome $\Gamma$
then it is $(\rho, \delta)$-robust with respect to $\Gamma$ at time $t$ where 
$\delta = F^\Gamma(\tilde{\mathcal{C}}^{+\rho}, t) - F^\Gamma(\tilde{\mathcal{C}}^{-\rho}, t) = 
F^\Gamma(\mathcal{C}, t/(1-\rho)) - F^\Gamma(\mathcal{C}, t/(1+\rho))$.
\end{corollary}

For many SSA processes, it may not be obvious whether they are monotonic.
We would like a simple ``syntactic" property of the SCRN that guarantees monotonicity and can be easily checked.
The following lemma makes it easy to prove monotonicity in some simple cases.
\begin{lemma}   \label{lem:syntacticmonotonicity}
Let $\mathcal{C}$ be an SSA process and $\Gamma$ an outcome of SCRN $\mathcal{S}$.
If every species is a reactant in at most one reaction in $\mathcal{S}$,
and there is a set $\{ n_j \}$ such that outcome $\Gamma$ occurs as soon as every reaction $R_j$ has fired at least $n_j$ times,
then $\mathcal{C}$ is monotonic with respect to $\Gamma$.
\end{lemma}
\begin{proof}
The restriction on $\Gamma$ allows us phrase everything in terms of counting reaction occurrences.
For every reaction $R_j$, define $F_j(n,t)$ to be the probability that $R_j$ has fired at least $n$ times within time $t$.
Now suppose we induce some reaction to fire by fiat.
The only way this can decrease some $F_j(n,t)$ is if it decreases the count of a reactant of $R_j$ or makes it more likely that some reaction $R_{j'}$ ($j' \neq j$) will decrease the count of a reactant of $R_j$.
Either possibility is avoided if the SCRN has the property that any species is a reactant in at most one reaction.
Since $F^\Gamma(\mathcal{C}, t) = \prod_j F_j(n_j,t)$, this quantity cannot decrease as well,
and monotonicity follows.
\end{proof}

\subsection{Robust Embedding of a TM in an SCRN}    \label{sec:robustembedding}

Since we are trying to bound how the complexity of the prediction problem scales with increasing bounds on $t$ and $C$ but not with different SCRNs, 
we need a method of embedding a TM in an SCRN in which the SCRN is independent of the input length.
Among such methods available~\citep{angluin06,shukipaper}, asymptotically the most efficient and therefore best for our purposes is the construction of Angluin et al.
This result is stated in the language of distributed multi-agent systems rather than molecular systems;
the system is a well-mixed set of ``agents" that randomly collide and exchange information.
Each agent has a finite state.
Agents correspond to molecules (the system preserves a constant molecular count $m$);
states of agents correspond to the species,
and interactions between agents correspond to reactions in which both molecules are potentially transformed.

Now for the details of the SCRN implementation of Angluin's protocol.
Suppose we construct an SCRN corresponding to the Angluin et al system as follows:
Agent states correspond to species (i.e., for every agent state $i$ there is a unique species $S_i$). 
For every pair of species $S_{i_1}, S_{i_2}$, $(i_1 \leq i_2)$ we add reaction $S_{i_1} + S_{i_2} \rightarrow S_{i_3} + S_{i_4}$ if the population protocol transition function specifies $(i_1, i_2) \mapsto (i_3, i_4)$.
Note that we allow null reactions of the form $S_{i_1} + S_{i_2} \rightarrow S_{i_1} + S_{i_2}$ including for $i_1 = i_2$.
For every reaction $R_j$, we'll use rate constant $k_j = 1$.
The sum of all reaction propensities is $\lambda = \frac{m(m-1)}{2V}$ since every molecule can react with any other molecule.\footnote{Just to confirm, splitting the reactions between the same species and between different species, the sum of the propensities is
$\sum_i \frac{x_i(x_i-1)}{2V} + \sum_{i < i'} \frac{x_i x_{i'}}{V} = 
\frac{1}{2V}(\sum_{i}x_i x_{i} - \sum_i x_i + 2 \sum_{i<i'} x_i x_{i'}) =
\frac{1}{2V}(\sum_{i,i'}x_i x_{i'} - \sum_i x_i) = \frac{n(n-1)}{2V}$ using the fact that $2 \sum_{i<i'} x_i x_{i'} = \sum_{i \neq i'} x_i x_{i'}$ and $\sum_i x_i x_i + \sum_{i \neq i'} x_i x_{i'} = \sum_{i,i'}x_i x_{i'}$.}
The time until next reaction is an exponential random variable with rate $\lambda$.
Note that the transition probabilities between SCRN states are the same as the transition probabilities between the corresponding configurations in the population protocol since in the SCRN every two molecules are equally likely to react next.
%XXX Maybe explicitly prove this here.
Thus the SSA process is just a continuous time version of the population protocol process (where unit ``time" expires between transitions).
Therefore the SCRN can simulate a TM in the same way as the population protocol.

But first we need to see how does time measured in the number of interactions correspond to the real-valued time in the language of SCRNs?

\begin{lemma}   \label{lem:chernoff}
If the time between population protocol interactions is an exponential random variable with rate $\lambda$, then for any positive constants $c, c_1, c_2$ such that $c_1 < 1 < c_2$, there is $N_0$ such that for all $N > N_0$, 
$N$ interactions occur between time $c_1 N /\lambda$ and $c_2 N /\lambda$ with probability at least $1-N^{-c}$.
\end{lemma}
\begin{proof}
The Chernoff bound for the left tail of a gamma random variable $T$ with shape parameter $N$ and rate $\lambda$ is 
$\Pr[T \leq t] \leq (\frac{\lambda t}{N})^N e^{N-\lambda t}$ for $t < N/\lambda$.
Thus $\Pr[T \leq c_1 N/\lambda] \leq ({c_1} e^{1-c_1})^N$.
Since $c_1 e^{1-c_1} < 1$ when $c_1 \neq 1$, 
$\Pr[T \leq c_1 N/\lambda] < N^{-c}$ for large enough $N$.
An identical argument applies to the right tail Chernoff bound $\Pr[T \geq t] \leq (\frac{\lambda t}{N})^N e^{N-\lambda t}$ for $t > N/\lambda$.
\end{proof}

The following lemma reiterates that an arbitrary computational problem can be embedded in a chemical system, 
and also shows that the chemical computation is robust with respect to the outcome of the computation.
For a given TM and agent count $m$, let $\vec{x_{f_0}} $ and $\vec{x_{f_1}} $ be SCRN states corresponding to the TM halting with a $0$ and $1$ output respectively.

\begin{lemma}    \label{lem:TMsim}
Fix a perturbation bound $\rho > 0$, $\delta > 0$, and a randomized TM $M$ with a Boolean output.
There is an SCRN implementing Angluin et al's population protocol, 
such that if $M(x)$ halts in no more than $t_{tm}$ steps using no more than $s_{tm}$ time, then
starting with the encoding of $x$ and using $m = O(1) 2^{s_{tm}}$ molecules,
at any time $t \geq t_{ssa} = O(1) V t_{tm} \log^4 (m) /m$ the SSA process is in $\vec{x_{f_b}} $ with probability that is within $\delta$ of the probability that $M(x) = b$.
Further, this SSA process is $(\rho, \delta)$-robust with respect to states $\vec{x_{f_0}}$ and $\vec{x_{f_1}}$ at all times $t \geq t_{ssa}$.
\end{lemma}

The first part of the lemma states that we can embed an arbitrary TM computation in an SCRN,
such that the TM computation is performed fast and correctly with arbitrarily high probability.
The second part states that this method can be made arbitrarily robust to perturbations of reaction propensities.
The first part follows directly from the results of~\citet{angluin06}, while
the second part requires some additional arguments on our part.

If we only wanted to prove the first part,
fix any randomized TM $M$ with a Boolean output and any constant $\delta > 0$.
There is a population protocol of Angluin et al that can simulate the TM's computation on arbitrary inputs as follows:
If on some input $x$, $M$ uses $t_{tm}$ computational time and $s_{tm}$ space,
then the protocol uses $m=O(1) 2^{s_{tm}}$ agents, 
and the probability that the simulation is incorrect or takes longer than $N = O(1) t_{tm} m \log^4{m}$ interactions is at most $\delta/2$.
This is proved by using Theorem 11 of~\citet{angluin06}, combined with the standard way of simulating a TM by a register machine using multiplication by a constant and division by a constant with remainder.
The total probability of the computation being incorrect or lasting more than $N$ interactions obtained is at most $O(1) t_{tm} m^{-c}$.
Since for any algorithm terminating in $t_{tm}$ steps,
$2^{s_{tm}} \geq O(1) \, t_{tm}$,
we can make sure this probability is at most $\delta/2$ by using a large enough constant in $m = O(1) 2^{s_{tm}}$.
%XXX: Reminder: verify why their "log-space" construction seems more complicated than necessary.
By Lemma~\ref{lem:chernoff}, 
the probability that $O(1) N$ interactions take longer than $O(1)N/\lambda$ time to occur is at most $\delta/2$.
Thus the total probability of incorrectly simulating $M$ on $x$ or taking longer than $O(1) N/\lambda = O(1) V t_{tm} \log^4(m)/m$ time is at most $\delta$. 
The Boolean output of $M$ is indicated by whether we end up in state $\vec{x_{f_0}}$ or $\vec{x_{f_0}}$.
(If the computation was incorrect or took too long we can be in neither.)
This proves the first part of the lemma.

We now sketch out the proof of how the robustness of the Angluin et al system can be established,
completing the proof of Lemma~\ref{lem:TMsim}.
The whole proof requires retracing the argument in the original paper; 
here, we just outline how this retracing can be done.
First, we convert the key lemmas of their paper to use real-valued SCRN time rather than the number of interactions.
The consequences of the lemmas (e.g.,\ that something happens before something else) are preserved and thus the lemmas can be still be used as in the original paper to prove the corresponding result for SCRNs.
The monotonicity of the processes analyzed in the key lemmas can be used to argue that the overall construction is robust.

We need the following consequence of Lemma~\ref{lem:monotonicity}:

\begin{corollary}  \label{cor:between}
If an SSA process $\mathcal{C}$ is monotonic for outcome $\Gamma$,
and with probability $p$ it enters $\Gamma$ after time $t_1$ but before time $t_2$,
then for any $\rho$-perturbation $\tilde\mathcal{C}$ of $\mathcal{C}$, the probability of entering $\Gamma$ after time $t_1/(1+\rho)$ but before time $t_2/(1-\rho)$ is at least $p$.
\end{corollary}

\begin{proof}
Let $p_1 = F^\Gamma(\mathcal{C}, t_1)$ and $p_2 = F^\Gamma(\mathcal{C}, t_2)$.
Using Lemma~\ref{lem:monotonicity} we know that $\forall t$, 
$F^\Gamma(\mathcal{C}, t/(1-\rho))      \geq   F^\Gamma(\tilde\mathcal{C}, t)$.
Thus, $p_1 = F^\Gamma(\mathcal{C}, t_1)      \geq   F^\Gamma(\tilde\mathcal{C}, (1-\rho) t_1)$.
Similarly we obtain 
$p_2 = F^\Gamma(\mathcal{C}, t_2)      \leq   F^\Gamma(\tilde\mathcal{C}, (1+\rho) t_2)$.
Thus $F^\Gamma(\tilde\mathcal{C}, (1+\rho) t_2) - F^\Gamma(\tilde\mathcal{C}, (1-\rho) t_1) \geq p_2 - p_1 = p$.
\end{proof}

As an example let us illustrate the conversion of Lemma~2 of~\citet{angluin06}.
The Lemma bounds the number of interactions to infect $k$ agents in a ``one-way epidemic" starting with a single infected agent.
In the one-way epidemic, a non-infected agent becomes infected when it interacts with a previously infected agent.
With our notation, this lemma states: 
\begin{quote}
Let $N(k)$ be the number of interactions before a one-way epidemic starting with a single infected agent infects $k$ agents.
Then for any fixed $\eps > 0$ and $c > 0$, there exist positive constants $c_1$ and $c_2$ such that for sufficiently large total agent count $m$ and any $k > m^\eps$, $c_1 m \ln k \leq N(k) \leq c_2 m \ln k$ with probability at least $1- m^{-c}$.
\end{quote}
For any $m$ and $k$ we consider the corresponding SSA process $\mathcal{C}$ and outcome $\Gamma$ in which at least $k$ agents are infected.
Since the bounds on $N(k)$ scale at least linearly with $m$, we can use Lemma~\ref{lem:chernoff} to obtain: 
\begin{quote}
Let $t(k)$ be the time before a one-way epidemic starting with a single infected agent infects $k$ agents.
Then for any fixed $\eps > 0$ and $c > 0$, there exist positive constants $c_1$ and $c_2$ such that for sufficiently large total agent count $m$ and any $k > m^\eps$, $c_1 m \ln (k)/\lambda \leq t(k) \leq c_2 m \ln (k)/\lambda$ with probability at least $1- m^{-c}$.
\end{quote}
Finally consider the SSA process of the one-way epidemic spreading.
The possible reactions either do nothing (reactants are either both infected or both non-infected),
or a new agent becomes infected.
It is clear that for any $m$ and $k$, $\mathcal{C}$ is monotonic with respect to outcome $\Gamma$ in which at least $k$ agents are infected.
This allows us to use Corollary~\ref{cor:between} to obtain:
\begin{quote}
Fix any $\rho > 0$,
and let $t(k)$ be the time before a one-way epidemic starting with a single infected agent infects $k$ agents in some corresponding $\rho$-perturbation.
Then for any fixed $\eps > 0$, $c > 0$, there exist positive constants $c_1$ and $c_2$ such that for sufficiently large total agent count $m$ and any $k > m^\eps$, $c_1 m \ln (k)/(\lambda (1+\rho)) \leq t(k) \leq c_2 m \ln (k)/(\lambda (1-\rho))$ with probability at least $1- m^{-c}$.
\end{quote}
Since $\rho$ is a constant, what we have effectively done is convert the result in terms of interactions to a result in terms of real-valued time that is robust to $\rho$-perturbations simply by dividing by $\lambda$ and using different multiplicative constants.

The same process can be followed for the key lemmas of Angluin et al (Lemma~3 through Lemma~8).
This allows us to prove a robust version of Theorem~11 of Angluin et al by retracing the argument of their paper using the converted lemmas and the real-valued notion of time throughout.
Since the only way that time is used is to argue that something occurs before something else, 
the new notion of time, obtained by dividing by $\lambda$ with different constants, 
can always be used in place of the number of interactions.

\subsection{Proof of Theorem~\ref{thm:optimality}: Lower Bound on the Computational Complexity of the Prediction Problem}       \label{sec:optimality}
In this section we prove Theorem~\ref{thm:optimality} from the text which lower-bounds the computational complexity of the prediction problem
as a function of $m$, $t$, and $C$.
The bound holds even for arbitrarily robust SSA processes.
The theorem shows that this computational complexity is at least linear in $t$ and $C$,
as long as the dependence on $m$ is at most polylogarithmic.
The result is a consequence of the robust embedding of a TM in an SCRN (Lemma~\ref{lem:TMsim}).

Let the prediction problem be specified by giving the SSA process (via the initial state and volume),
the target time $t$, and
the target outcome $\Gamma$ in some standard encoding such that whether a state belongs to $\Gamma$ can be computed in time polylogarithmic in $m$.
\begin{theorem*}
Fix any perturbation bound $\rho > 0$ and $\delta > 0$.
Assuming the hierarchy conjecture (Conjecture~\ref{con:hierarchyconjecture}),
there is an SCRN $\mathcal{S}$ such that
for any prediction algorithm $\mathcal{A}$ and constants $c_1, c_2, \beta, \eta, \gamma >0$,
there is an SSA process $\mathcal{C}$ of $\mathcal{S}$ and a 
$(m, t, C, 1/3)$-prediction problem $\mathcal{P}$ of $\mathcal{C}$ 
such that $\mathcal{A}$ cannot solve $\mathcal{P}$ in computational time 
$c_1\, (\log m)^{\beta} \, t^\eta \, (C+c_2)^\gamma$ if $\eta < 1$ or  $\gamma < 1$.
Further, $\mathcal{C}$ is $(\rho, \delta)$-robust with respect to $\mathcal{P}$.
\end{theorem*}

Suppose someone claims that for any fixed SCRN, 
they can produce an algorithm for solving $(m, t, C, 1/3)$-prediction problems for SSA processes of this SCRN assuming the SSA process is $(\rho, \delta)$-robust with respect to the prediction problem for some fixed $\rho$ and $\delta$,
and further they claim the algorithm runs in computation time at most
\begin{equation}
O(1)\, (\log(m))^\beta \, t^\eta \, (C+O(1))^\gamma
\label{eqn:claimedalg}
\end{equation}
for some $\eta < 1$ ($\beta, \gamma > 0$).
We argue that assuming the hierarchy conjecture is true, such a value of $\eta$ is impossible.

To achieve a contradiction of the hierarchy conjecture,
consider any function probabilistically computable in $t_{tm}(n) = O(1) n^\zeta$ time and $s_{tm}(n) = O(1) n$ space for $\zeta = \frac{\beta + 4\eta}{1-\eta} + 1$.
Construct a randomized TM having error at most $1/24$ by running the original randomized TM $O(1)$ times and taking the majority vote.
Use Lemma~\ref{lem:TMsim} to encode the TM probabilistically computing this function in a $(\rho,\delta)$-robust SSA process such that the error of the TM simulation is at most $1/24$.
Then predicting whether the process ends up in state $\vec{x_{f_0}}$ or $\vec{x_{f_1}}$ provides a probabilistic algorithm for computing this function.
The resulting error is at most $1/24 + 1/24 + 1/3 = 5/12 < 1/2$, where the first term $1/24$ is the error of the TM, the second term $1/24$ is for the additional error of the TM embedding in the SSA process, and the last term $1/3$ is for the allowed error of the prediction problem. 
By repeating $O(1)$ times and taking the majority vote, this error can be reduced below $1/3$, thereby satisfying the definition of probabilistic computation.
How long does this method take to evaluate the function?
We use $V = m$ so that $C$ is a constant,
resulting in $t_{ssa} = O(1) t_{tm}(n) \log^4{m} = O(1) n^{\zeta+4}$ since $m = O(1) 2^n$.
Setting up the prediction problem by specifying the SSA process (via the initial state and volume),
target final state and time $t_{ssa}$
requires $O(1) \log{m} = O(1) \, n$ time.\footnote{By the construction of \citet{angluin06}, setting up the initial state requires letting the binary expansion of the molecular count of a certain species be equal the input. 
Since the input is given in binary and all molecular counts are represented in binary, this is a linear time operation.
Setting up the final state $\vec{x_{f_0}}$ or $\vec{x_{f_1}}$ is also linear time.
Computing the target time for the prediction problem $t_{ssa}$ is asymptotically negligible.}
Then the prediction problem is solved in computation time 
$O(1)(\log(m))^\beta t_{ssa}^\eta = O(1) n^{\beta + (\zeta + 4)\eta}$. 
Thus the total computation time is $O(1) (n^{\beta + (\zeta + 4)\eta} + n)$ which, by our choice of $\zeta$, is less than $O(1) n^\zeta$,
leading to a contradiction of the hierarchy conjecture.

Is $\gamma < 1$ possible?
Observe that if $\gamma < \eta$ then the claimed running time of the algorithm solving the prediction problem (expression~\ref{eqn:claimedalg}) with time $t_{ssa} = O(1) V t_{tm}(n) \log^4(m)/m$ can be made arbitrarily small by decreasing $V$.
This leads to contradiction of the hierarchy conjecture.
Therefore $\gamma \geq \eta \geq 1$.

\subsection{On Implementing BTL on a Randomized TM}      \label{sec:TMimplementation}
The idealized BTL algorithm presented in Section~\ref{sec:algorithm} relies on infinite precision real-value arithmetic,
while only finite precision floating-point arithmetic is possible on a TM.
Further, the basic randomness generating operation available to a randomized TM is choosing one of a fixed number of alternatives uniformly, 
which forces gamma and binomial draws to be approximated.
This complicates estimates of the computation time required per leap, and 
also requires us to ensure that we can ignore round-off errors in floating-point operations and tolerate approximate sampling in random number draws.

Can we implement gamma and binomial random number generators on a randomized TM and how much computational time do they require?
It is easy to see that arbitrary precision uniform $[0,1]$ random variates can be drawn on a randomized TM in time linear in precision.
It is likely that approximate gamma and binomial random variables can be drawn using methods available in the numerical algorithms literature which uses uniform variate draws as the essential primitive.
Since many existing methods for efficiently drawing (approximate) gamma and binomial random variables involve the rejection method, the computation time for these operations is likely to be an expectation.
Specifically, it seems reasonable that drawing gamma and binomial random variables can be approximately implemented on a randomized TM such that the expected time of these operations is polynomial in the length of the floating-point representation of the distribution parameters and the resultant random quantity.%
\footnote{The numerical algorithms literature, which assumes that basic floating point operations take unit time,
describes a number of algorithms for drawing from an (approximate) standard gamma distribution~\citep{ahrens1978ggv},
and from a binomial distribution~\citep{kachitvichyanukul1988brv}, such that the expected number of floating-point operations does not grow as a function of distribution parameters (however, some restrictions on the parameters may be required).
On a TM basic arithmetic operations take polynomial time in the length of the starting numerical values and the calculated result.}

The computational complexity of manipulating integer molecular counts on a TM is polylogarithmic in $m$.
Let $l$ be an upper bound on the expected computational time required for drawing the random variables and real number arithmetic;
$l$ is potentially a function of $m$, $V$, $t$, and the bits of precision used.
Using Markov's inequality and Theorem~\ref{thm:algspeed} we can then obtain a bound on the total computation time that is true with arbitrarily high probability.
We make the TM keep track of the total number of computational steps it has taken\footnote{Compute the bound and write this many $1$'s on a work tape, and after each computational step, count off one of the $1$'s until no more are left.} and cut off computation when it exceeds the expectation by some fixed factor.
Then we obtain the following bound on the total computation time:
$O(1)((\log(m))^{O(1)} +l) \, t \, (C+ O(1))$.

We have three sources of error.
First, since BTL simulates a $\rho$-perturbation rather than the original SSA process, the probability of the outcome may be off by $\delta_1$, assuming the SSA process was $(\rho, \delta_1)$-robust.
Further, since we are using finite precision arithmetic and only approximate random number generation,
the deviation from the correct probability of the outcome may increase by another $\delta_2$.
Finally, there is a $\delta_3$ probability that the algorithm cuts off computation before it completes.
We want to guarantee that the total error $\delta_1 + \delta_2 + \delta_3 \leq \delta$, 
fulfilling the requirements of solving the $(m,t,C,\delta)$-prediction problem.
While $\delta_1 < \delta$ as a precondition, 
and we can make $\delta_3$ arbitrarily small,
a rigorous bound on $\delta_2$ is beyond the scope of this paper.\footnote{We conjecture that for any fixed $\delta_2$,
we can find some fixed amount of numerical precision to not exceed $\delta_2$
for $(\rho, \delta_1)$-robust processes.
We would like to show that robustness according to our definition implies robustness to round-off errors and approximate random number generation.
While this conjecture has strong intuitive appeal, it seems difficult to prove formally,
and represents an area for further study.
%XXX: Note that it is easy to prove that deviations in the time at which we make the prediction caused by round-off errors don't matter for robust processes.
}

\vspace{0.3in}
\textbf{Acknowledgments:}
I thank Erik Winfree and Matthew Cook for providing invaluable support, technical insight, corrections and suggestions.
This work was supported by NSF Grant No. 0523761 to Winfree and NIMH Training Grant MH19138-15 to CNS.

\end{document}